\documentclass[%
 reprint,
 amsmath,amssymb,
aps,
]{revtex4-1}

\usepackage{graphicx}
\usepackage{dcolumn}
\usepackage{bm}
\usepackage[justification=justified, format=plain]{caption}
\usepackage{subcaption}
\usepackage{physics}
\usepackage{mathtools}

\usepackage[unicode=true,bookmarks=false,breaklinks=false,pdfborder={0 0 1},colorlinks=true]
 {hyperref}
\hypersetup{
 citecolor=blue,linkcolor=blue,urlcolor=blue}

\begin{document}

\preprint{MANUSCRIPT VERSION}

\title{Two-photon spontaneous emission of an atom in a cosmic string background}

\author{Lucas Weitzel} 
 \email{weitzel@pos.if.ufrj.br}
\author{Y. Muniz}%
 \email{yurimuniz7@gmail.com}
\author{C. Farina}
\email{farina@if.ufrj.br}
\author{Carlos A. D. Zarro} 
 \email{carlos.zarro@if.ufrj.br}
\affiliation{Instituto de F\'{i}sica, Universidade Federal do Rio de Janeiro, Avenida Athos da Silveira Ramos, 149, Cidade Universit\'{a}ria, Rio de Janeiro-RJ, Brazil}

\date{\today}

\begin{abstract}
    
    It is well known that the vicinities of an atomic system may substantially affect its radiative properties. In this work,  we consider the influence of a cosmic string background in the spontaneous emission  of an excited atom.  We start by computing the one-photon spontaneous emission rate of a quantum emitter, which is a narrow band process, and then we analyze the more complex case of the two-photon spontaneous emission, which is a broad band and much richer phenomenon.  In the former case, we analyze not only the behavior of  the decay rate with the distance from the atom to the string, but also with the deficit angle associated with the cosmic string metric.  In the latter case, we show that the spectral distribution of the emitted photons is substantially affected by the cosmic string background.
\end{abstract}

\maketitle

\section{\label{sec:intro}Introduction}

The spontaneous decay of a quantum emitter is a fundamental phenomenon of physics and it is responsible for most of the light we observe \cite{milonni1984}. Since a substantial amount of information we gather from the universe components comes from their emission spectrum, spontaneous emission (SE) also plays a key role in astronomy and cosmology.  For example, the so called 21cm line, which is very important in radio astronomy and cosmology, has its origin in the transition between two hyperfine levels of the ground state of hydrogen \cite{purcell1951,furlanetto2006,pritchard2012}, which partially comes from SE. In general, an excited atom decays by emitting a single photon; however, higher order decay pathways such as two-photon spontaneous emission (TPSE) exist and may not be negligible depending on the system features. An isolated hydrogen atom, for instance, in the $2s$ metastable state, cannot decay by one-photon spontaneous emission due to selection rules. In this case, TPSE is the fastest pathway to the ground state and dominates the $2s \rightarrow 1s$ transition, despite being a billion times slower than conventional one-photon emission (e.g. $2p \rightarrow 1s$ transition in hydrogen) \cite{breit1940}. Furthermore, the $2s \rightarrow 1s$ transition in hydrogen and ionized helium is the fundamental phenomenon behind the emission spectrum of planetary nebula \cite{spitzer1951,gurzadyan2013} and microwave cosmological background generated during the recombination period \cite{wong2006,hirata2008,chluba2011}. Besides its importance in cosmology, TPSE processes have been intensely studied in other scenarios since its theoretical prediction by G\"oppert-Meyer (particularly, see Ref. \cite{ilakovac2006two} references therein). 

It is well known that any changes in the electromagnetic vacuum mode field influence the radiative properties of an emitter, particularly its SE rate. This phenomenon is known as the Purcell effect \cite{purcell1946} and can be naturally achieved by the presence of material bodies in the vicinities of the emitter. The Purcell effect occurs not only in the one-photon SE process, in which it is widely studied \cite{Haroche, Lodahl}, but also in the TPSE. For instance, it has been shown that the TPSE can be orders of magnitude larger if the emitter is placed near polar dielectrics \cite{rivera2017making}, graphene monolayers \cite{rivera2016shrinking}, and atomically thin plasmonic nanostructures \cite{muniz2020two}. Nevertheless, environment deviations from Minkowski spacetime due to a gravitational  field  also  affect  the  vacuum  and  the  SE  of  an atom \cite{Bekenstein:1977mv}.  Recently the Purcell effect for one-photon SE has been studied in a cosmic string background \cite{Cai:2015ioa}. However, as far as the authors know, the influence of a gravitational field in a TPSE process has never been investigated.

One of the most important breakthroughs in last century physics was the concept of the spontaneous symmetry breaking. From this issue, it is possible to have two or more different vacua which are not equivalent, \textit{i.e.} vacua that cannot be transformed into each other by a continuous change of an arbitrary parameter. Between regions with different vacua, the so-called topological defects appear, which are characterized by non-trivial homotopy groups. Depending on these groups, the defects can be classified as magnetic monopoles, zero-dimensional; cosmic strings, one-dimensional; domain walls, two-dimensional; among others. An interested reader is referred to \cite{Mermin:1979zz,vilenkin2000cosmic} for a general discussion of the classification of topological defects.  Although they are very ubiquitous in condensed matter physics, from ferromagnetism and crystalline defects to vortexes in superfluid materials, in high energy physics and cosmology the interest rests upon the formation of defects during the early universe, where presumably there was a spontaneous symmetry breaking of a larger symmetry group, described in a grand-unified theory of particle physics. The idea was that a cosmic string could act like seeds for the formation of complex cosmological structures like galaxies, although recent data rules out at least a prominent role in the formation of these structures \cite{vilenkin2000cosmic,Vachaspati:2015cma}.  However there has been an increasing interest in these structures as a source for stochastic gravitational wave background in the early universe, as recent data from the NANOGrav Collaboration may suggest \cite{ellis2020cosmic, blasi2020has, samanta2021gravitational}, as well as a gravitational analogue model for the geometry and the interaction of quantum dots near a crystalline defect called disclination \cite{Moraes:2000xa,Hu:2017ytf}.

From the gravitational point of view the first solution of the linearized Einstein equation for a straight string with linear mass density $\mu$ was obtained in \cite{Vilenkin:1981kz}. It was shown that the space-time around a cosmic string has a conical singularity with deficit angle $\delta \phi$ proportional to $\mu$. That is, the cosmic string background is a flat space-time  where the $\phi$ variable has periodicity $2\pi -\delta \phi$, instead of $2\pi$ of the flat geometry. However, a more general solution for a cosmic string can be obtained through the solution of the Einstein equations \cite{Hiscock:1985uc}. This case can model a more realistic string with a finite thickness. Nonetheless the main features of the linearized solution are not altered: the cosmic string space-time presents an asymptotic conical geometry with the same deficit angle as before. Hence, for brevity, from now on we are going to consider only the linearized solution. 

The cosmic string space-time background causes some interesting effects. As the space-time is flat, a particle near the string is not gravitationally attracted by it. Moreover, due to the non-trivial topology induced by a cosmic string, two parallel light rays can be deflected by it, generating a gravitational lensing \cite{vilenkin1984cosmic}, and a charge near the string feels a self electric force \cite{linet}. Other features, such as the Aharonov-Bohm analog,  can be found in Ref. \cite{vilenkin2000cosmic, aliev1989gravitational}.

In this work, one- and two-photon spontaneous emission rates for an excited atom, at rest in the presence of cosmic string are discussed. The vacuum electromagnetic field modes will be modified by the new background, and, as a consequence, the above mentioned spontaneous emission rates will also be altered. We then compare these two modified emission rates with those  obtained with the excited atom in the Minkowski space-time and show that they can be substantially different. We discuss the dependence  of these emission rates not only with the distance  between the atom and the cosmic string but also with the mass linear density of the string. 

This manuscript is organized as follows: in Sec. \ref{sec:onephotonSS} we investigate the one-photon spontaneous emission for the above set-up. In Sec. \ref{sec:twophotonSS},
we discuss the TPSE for the same set-up and present our main results. Sec. \ref{sec:conclusions} is left for our final remarks and conclusions. 
Two appendixes have also been included: in the first one we briefly review the electromagnetic field modes near a cosmic string, while in the second one, some mathematical details  concerning the one-photon SE rate are presented.

\section{\label{sec:onephotonSS} One-Photon Spontaneous Emission in a Cosmic String Background}

\begin{figure}[h]
    \centering
    \includegraphics[width=\columnwidth]{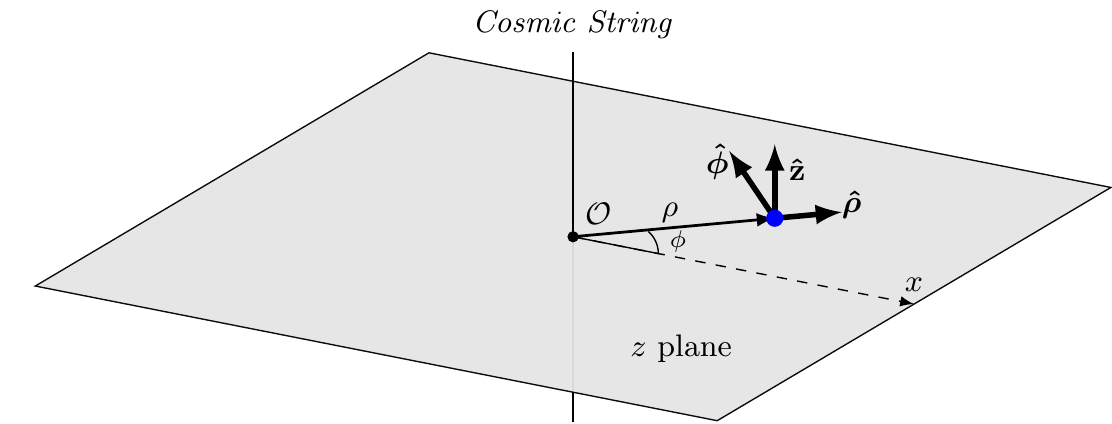}
    \caption{An excited atom (blue point) placed near a cosmic string. Using cylindrical coordinates, the atom is in the plane $z=0$ and its distance to the cosmic string is $\rho$. The orthonormal basis $\{\vu*\rho, \vu*\phi, \vu z\}$ is also depicted.}
    \label{fig:setup}
\end{figure}

Let us consider a quantum emitter in the vicinities of a straight and electrically neutral cosmic string of constant linear density of mass $\mu$.  We choose our axes so that the string is along the $\mathcal O z$ axis and the emitter is placed a distant $\rho$ from the string, as shown in Fig. \ref{fig:setup}.
The metric of this setup reads in cylindrical coordinates
    \begin{equation}
        \dd s^2=-c^{2}\,\dd t^2+\dd z^2+\dd r^2+\left(1-\frac{8G\mu}{c^2}\right)r^2\dd \phi^2,
    \end{equation}
where $G$ is the gravitational constant. This metric can be associated with a locally flat space-time, but with a global conical geometry, with a deficit angle $\delta\phi = 8\pi G\mu/c^2$. The mass linear density must satisfy the constraint $\mu<\frac{c^2}{4G}\approx 3.3\cdot 10^{26}$ kg/m, otherwise it would degenerate into another topology other than a conical one \cite{Hiscock:1985uc}.

This system can be described by a Hamiltonian given by $H=H_{A}+H_{F}+H_{\text{int}}$, where $H_{A}$, $H_F$ and $H_{\text{int}}$ are, respectively, the atomic, field and interaction Hamiltonians. In the Coulomb gauge and assuming that the dominant transition wavelengths are much greater than the quantum emitter dimensions, so that the dipole approximation is valid, the interaction Hamiltonian reads
\begin{align}
    H_{\text{int}}(\vb r)&=-\vb d\cdot \vb E(\vb r)\nonumber\\
    &=-\sum_{\alpha}\sqrt{\frac{\hbar \omega_\alpha}{2\epsilon_0}}\left[a_{\alpha}\vb d\cdot\vb A_{\alpha}(\vb r)-a^\dagger_{\alpha}\vb d\cdot\vb A^*_{\alpha}(\vb r)\right], \label{eq:Hint}
\end{align}
where $a^\dagger_{\alpha}$ and $a_{\alpha}$ stand for the creation and annihilation operators of a photon in mode $\alpha$, $\omega_\alpha$ is the photon frequency, $\vb d$ is the dipole moment operator of the quantum emitter which is placed at position $\vb{r}$, and $\{\vb A_{\alpha}\}$ is a complete set of solutions of the Helmholtz equation subjected to the boundary conditions imposed by the cosmic string as well as the Coulomb gauge restriction.

	\subsection{Methodology}

    The one-photon spontaneous emission (OPSE) rate can be obtained by using Fermi golden rule and first-order perturbation theory. In the initial state of our system, denoted by $\ket{e,0}$, the atom is in an excited state and the are no photons in the field; in the final state, denoted by $\ket{g, 1}$, the atom is in a state of lower energy (not necessarily its ground state) and there is one photon in the field in the mode $\alpha$. The OPSE rate can be written in terms of the field modes as \cite{Milonni}
	\begin{equation}
        \label{eq:Gamma2}
        \Gamma(\vb r) = \frac{\pi}{\epsilon_0\hbar}\sum_{\alpha}\omega_\alpha |\vb{d}_{eg}\cdot\vb{A}_{\alpha}(\vb{r})|^2\delta(\omega_\alpha - \omega_{eg}),
    \end{equation}
    where $\vb d_{eg}=\mel{e}{\vb d}{g}$ is the transition dipole moment and $\omega_{eg}$ is the transition frequency. From this expression and using the free-space electromagnetic field modes, namely $\vb A = e^{i\vb k\cdot \vb r}\vb e_{\vb k p}/\sqrt{V}$, it is possible to derive the corresponding free-space OPSE rate, \cite{Dirac:1927}
    \begin{equation}\label{eq:Gamma0}
        \Gamma_0=\frac{|\vb d_{eg}|^2\omega^2_{eg}}{3\pi\epsilon_0\hbar c^3}.
    \end{equation}
The influence of a cosmic string background on the OPSE rate can be obtained by inserting the corresponding field modes into Eq. \eqref{eq:Gamma2}. In appendix \ref{app:modes} we provide a brief derivation of these modes, which are given by
\begin{align}
    \vb{A}_{\vb k 0}&=\frac{\beta_{\vb k 0}c^2}{i\omega}\left(k_\perp^2\vu z + ik_z\grad_{\perp}\right)\nonumber \\
    &\times\left[J_{q|m|}\left(k_\perp\rho\right)e^{i(qm\phi+k_z z-\omega t)}\right], \label{eq:Amodo0}\\
    \vb{A}_{\vb k 1}&=-\beta_{\vb k 1} c\, \vu{z}\times\grad_{\perp}\left[J_{q|m|}\left(k_\perp\rho\right)e^{i(qm\phi+k_z z-\omega t)}\right], \label{eq:Amodo1}
\end{align}
where the indexes $0$ and $1$ indicate, respectively, the transverse magnetic and transverse electric modes of the field, 
$\{J_\nu(z)\}$ are the cylindrical Bessel functions, $q=2\pi/\phi_{0}$,
$m$ is an arbitrary integer,  $\phi_{0}=2\pi-\delta\phi$ and $\beta_{\vb k 0}$ and $\beta_{\vb k 1}$ are normalization constants such that $|\beta_{\vb k 0}|^2=|\beta_{\vb k 1}|^2=q/(2\pi k_\perp c)^2$.

It is convenient to calculate the OPSE rate in three different situations, namely, when the transition dipole moment of the quantum emitter is oriented along each of the cylindrical unit vectors, namely $\vb d_{eg}/|\vb d_{eg}|=\vu*\rho,\vu*\phi$ or $\vu z$. Substituting Eqs. \eqref{eq:Amodo0} and \eqref{eq:Amodo1} into Eq. \eqref{eq:Gamma2} we find (see Appendix \ref{app:OPSErates})
\begin{align}
    \label{eq:Gammaz}
    \frac{\Gamma_{\vu z}}{\Gamma_0}&=\frac{3q}{2}\sum_{m=-\infty}^\infty \int_0^1 \dd u\,\frac{u^3}{\sqrt{1-u^2}}J^2_{q|m|}\left(k_{eg}\rho u\right),\\
    \label{eq:Gammarho}
    \frac{\Gamma_{\vu*\rho}}{\Gamma_0}&=
    \frac{3q}{8}\sum_{m=-\infty}^\infty \int_0^{1} \dd u\,\frac{u}{\sqrt{1-u^2}}\nonumber\\& \times\left[\left(2-u^2\right)\left(J^2_{q|m|-1}\left(k_{eg}\rho u\right)+J^2_{q|m|+1}\left(k_{eg}\rho u\right)\right)\right.\nonumber\\&+2u^2J_{q|m|-1}\left(k_{eg}\rho u\right)J_{q|m|+1}\left(k_{eg}\rho u\right)\Big],\\
    \label{eq:Gammaphi}
    \frac{\Gamma_{\vu*\phi}}{\Gamma_0}&=\frac{3q}{8}\sum_{m=-\infty}^\infty \int_0^{1} \dd u\,\frac{u}{\sqrt{1-u^2}}\nonumber \\
    &\times\left[\left(2-u^2\right)\left(J^2_{q|m|-1}\left(k_{eg}\rho u\right)+J^2_{q|m|+1}\left(k_{eg}\rho u\right)\right)\right.\nonumber\\
    &-2u^2J_{q|m|-1}\left(k_{eg}\rho u\right)J_{q|m|+1}\left(k_{eg}\rho u\right)\Big].
\end{align}
The subscripts on the left hand sides of equations \eqref{eq:Gammaz}), \eqref{eq:Gammarho} and \eqref{eq:Gammaphi} indicate the direction of the emitter's transition dipole moment. These results are compatible with those found in Ref. \cite{Cai:2015ioa}. The SE rate for the isotropic case is simply given by $\Gamma=\frac13(\Gamma_{\vu z}+\Gamma_{\vu*\rho}+\Gamma_{\vu*\phi})$.

As a self-consistency test, let us re-obtain the OPSE rate in free-space  for a particular orientation of the transition dipole moment of the quantum emitter, say $ \frac{\Gamma_{\vu*\rho}}{\Gamma_0}$. This situation corresponds to take $\mu=0$ (or equivalently to take $q=1$)  in Eq. \eqref{eq:Gammarho}, namely,
\begin{align}
    \frac{\Gamma_{\vu*\rho}}{\Gamma_0}\Big|_{q=1}&=
    \frac{3}{8}\sum_{m=-\infty}^\infty \int_0^{1} \dd u\,\frac{u}{\sqrt{1-u^2}}\nonumber\\& \left[\left(2-u^2\right)\left(J^2_{|m|-1}\left(k_{eg}\rho u\right)+J^2_{|m|+1}\left(k_{eg}\rho u\right)\right)\right.\nonumber\\&+2u^2J_{|m|-1}\left(k_{eg}\rho u\right)J_{|m|+1}\left(k_{eg}\rho u\right)\Big].
\end{align}
Using the following properties of Bessel functions,
    \begin{equation}
    \sum_{m=-\infty}^{\infty}J_{|m|}^{2}(x)=J^2_0(x)+2\sum^\infty_{\nu=1} J^2_\nu(x)=1
    \end{equation}
    and
    \begin{equation}
        \sum_{m=-\infty}^{\infty}J_{|m|+1}(x)J_{|m|-1}(x)=0,
    \end{equation}
it is no difficult to see that Eq. \eqref{eq:Gamma0} is recovered, since
\begin{align}
    \frac{\Gamma_{\vu*\rho}}{\Gamma_0}&=
    \frac{3}{4}\sum_{m=-\infty}^\infty \int_0^{1} \dd u\,\frac{u\left(2-u^2\right)}{\sqrt{1-u^2}}=1\, .
\end{align}
Analogous calculations can be done for the other two orientations.

\subsection{Results and Discussions}

\begin{figure*}
     \centering
     \begin{subfigure}[b]{\columnwidth}
         \centering
        \includegraphics[width=\textwidth]{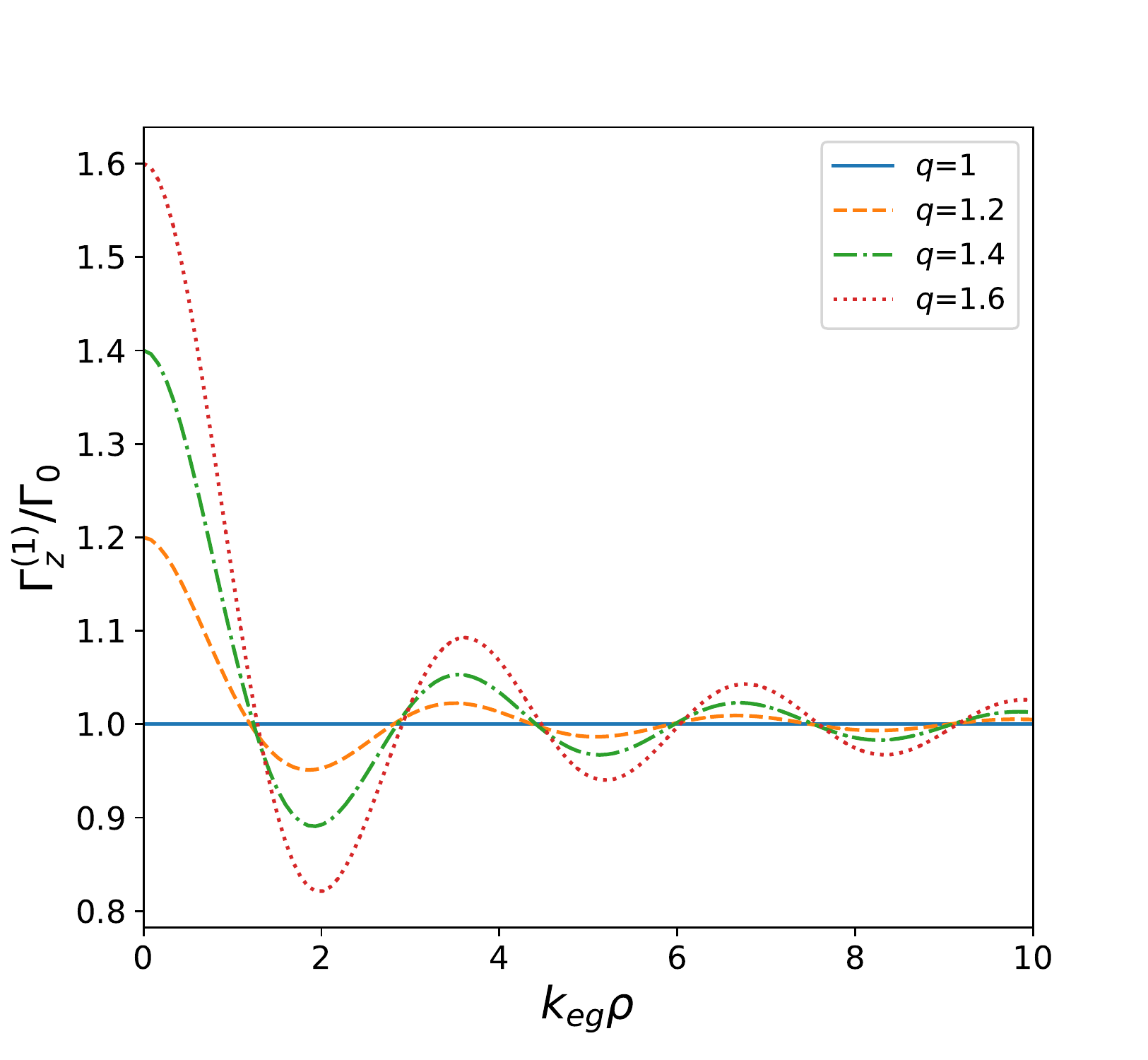}
        \caption{}
    \label{fig:Gammaz_r}
    \end{subfigure}
     \begin{subfigure}[b]{\columnwidth}
        \centering
        \includegraphics[width=\textwidth]{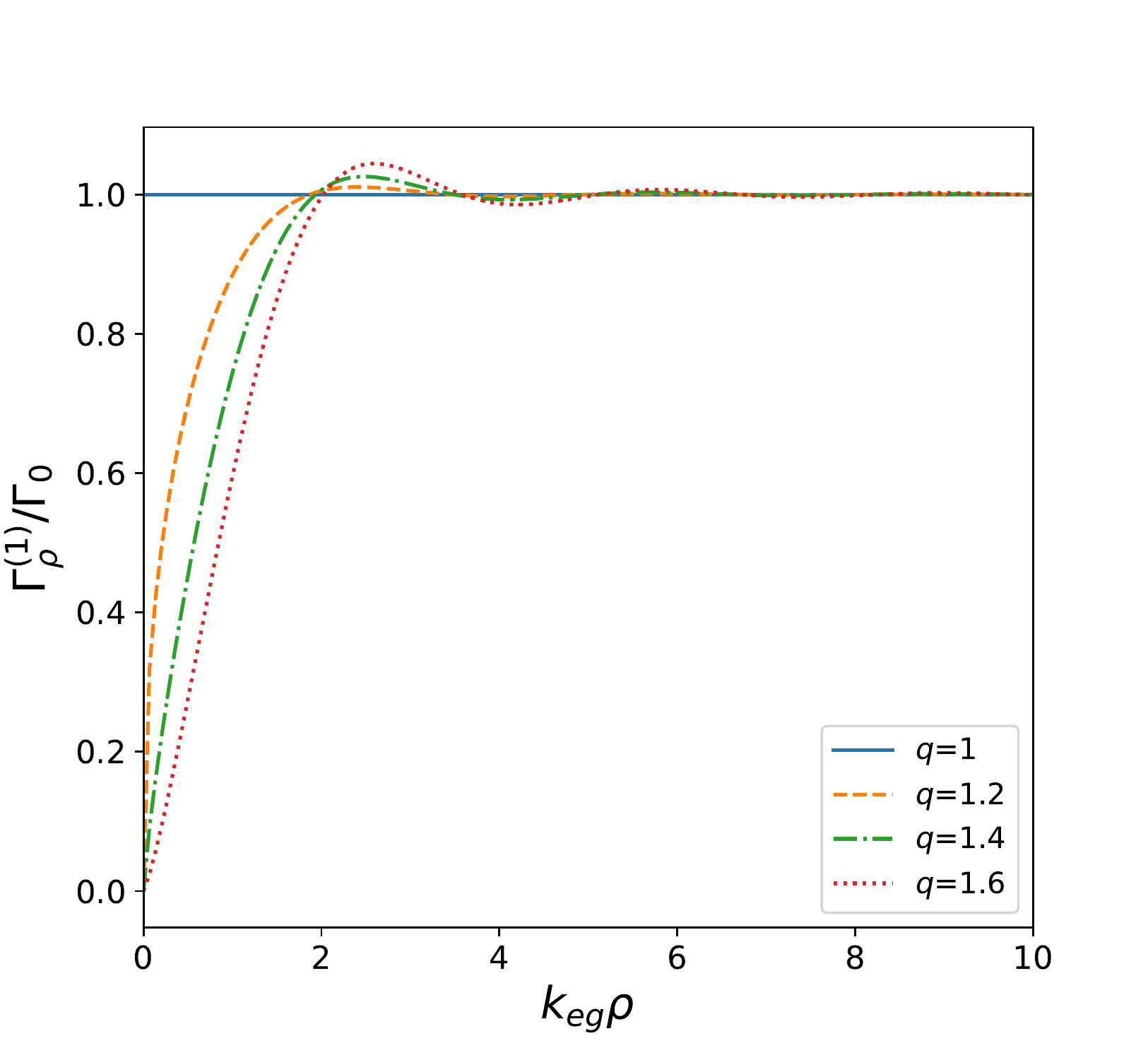}
        \caption{}
        \label{fig:Gammarho_r}
     \end{subfigure}
     \\
     \begin{subfigure}[b]{\columnwidth}
        \centering
        \includegraphics[width=\textwidth]{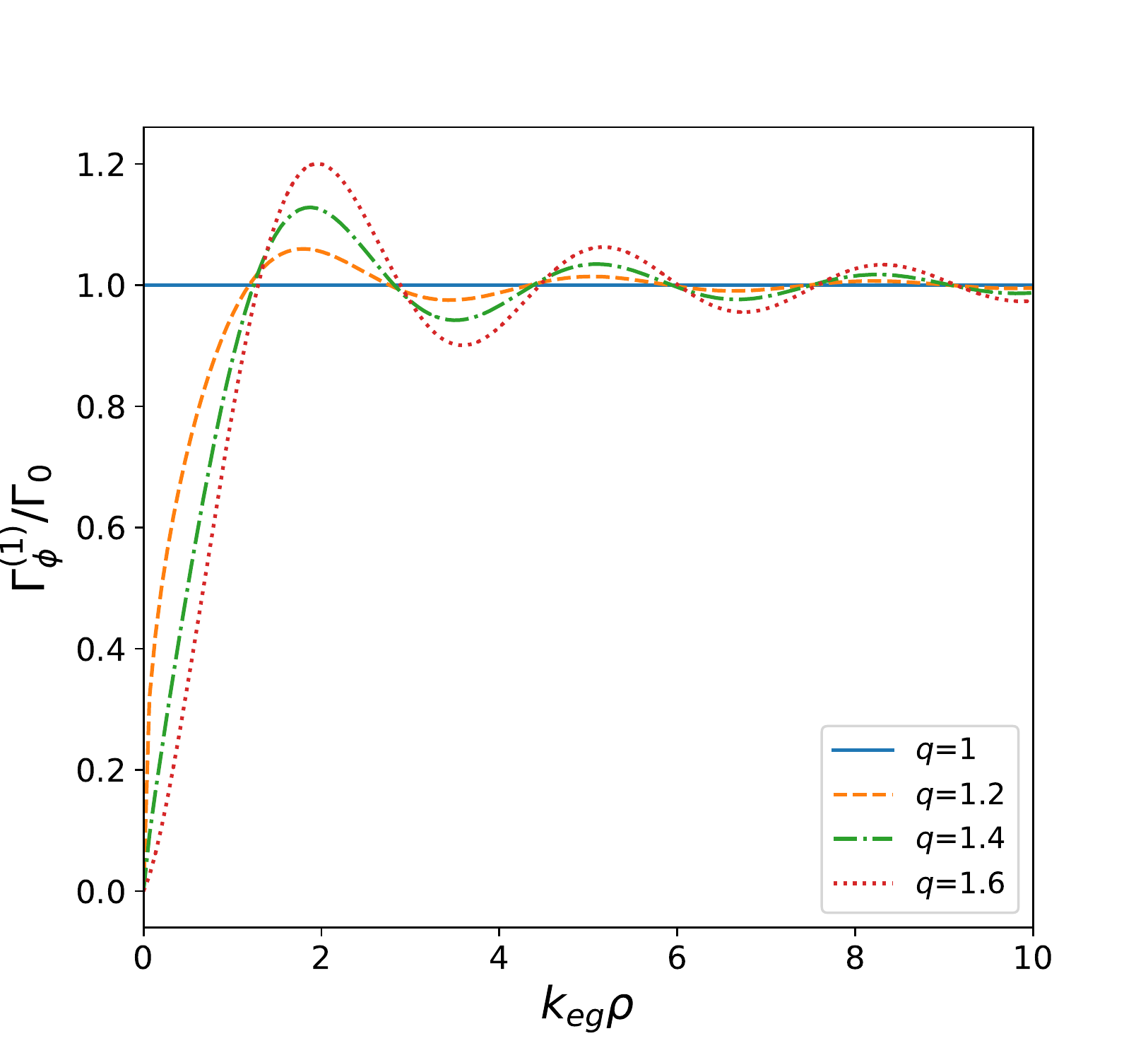}
        \caption{}
        \label{fig:Gammaphi_r}
     \end{subfigure}
     \begin{subfigure}[b]{\columnwidth}
        \centering
        \includegraphics[width=\textwidth]{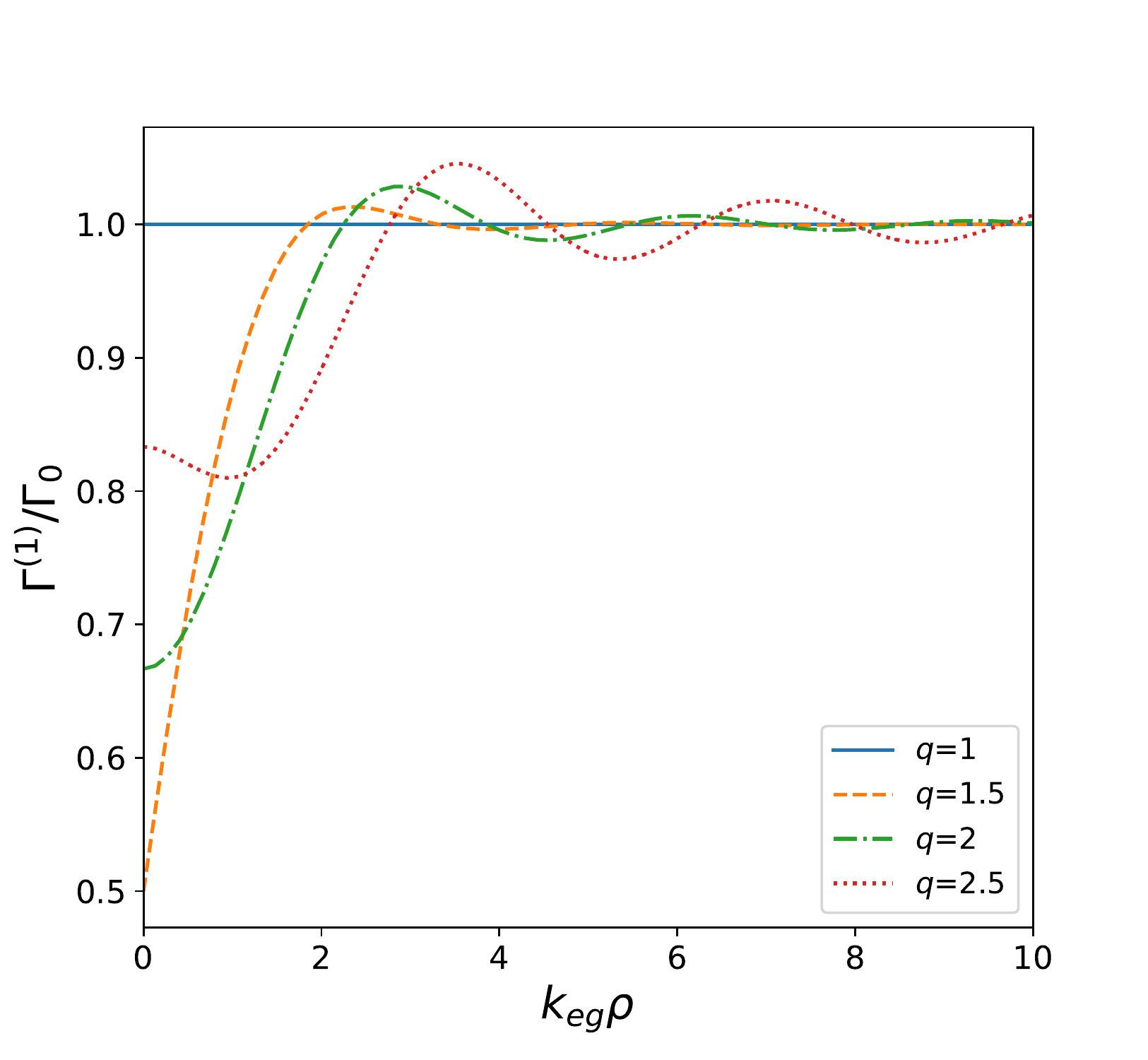}
        \caption{}
        \label{fig:Gammaiso_r}
    \end{subfigure}
    \caption{OPSE rates, as functions of the dimensionless distance $k_{eg}\rho$  (essentially, distance in units of the transition wavelength), for different values of parameter $q$, when the transition dipole moment is (\subref{fig:Gammaz_r}) parallel to the direction of the string; (\subref{fig:Gammarho_r}) perpendicular to the direction of the string but belongs to the plane containing the string and the atom;
    (\subref{fig:Gammaphi_r}) perpendicular to the direction of the string and also perpendicular to the plane containing the string and the atom;
     (\subref{fig:Gammaiso_r}) randomly oriented.}
    \label{fig:Gamma_r}
\end{figure*}

In Fig. \ref{fig:Gamma_r} we plot the normalized SE rates for each of the above mentioned orientations of the transition dipole moment, as well as for the isotropic case, as functions of distance $\rho$ from the atom to the string for different values of $q$. As a first self-consistency check, notice that for $k_{eg}\rho\gg 1$ we recover the SE rate in free-space, as expected, since the greater the distance between the atom and the string the smaller will be the influence of the string.  As already mentioned  in the previous subsection, the free-space result can also be achieved if we take the limit $q\to 1$, since such a limit means to remove the string. 

\begin{figure*}
     \centering
     \begin{subfigure}[b]{\columnwidth}
         \centering
        \includegraphics[width=\textwidth]{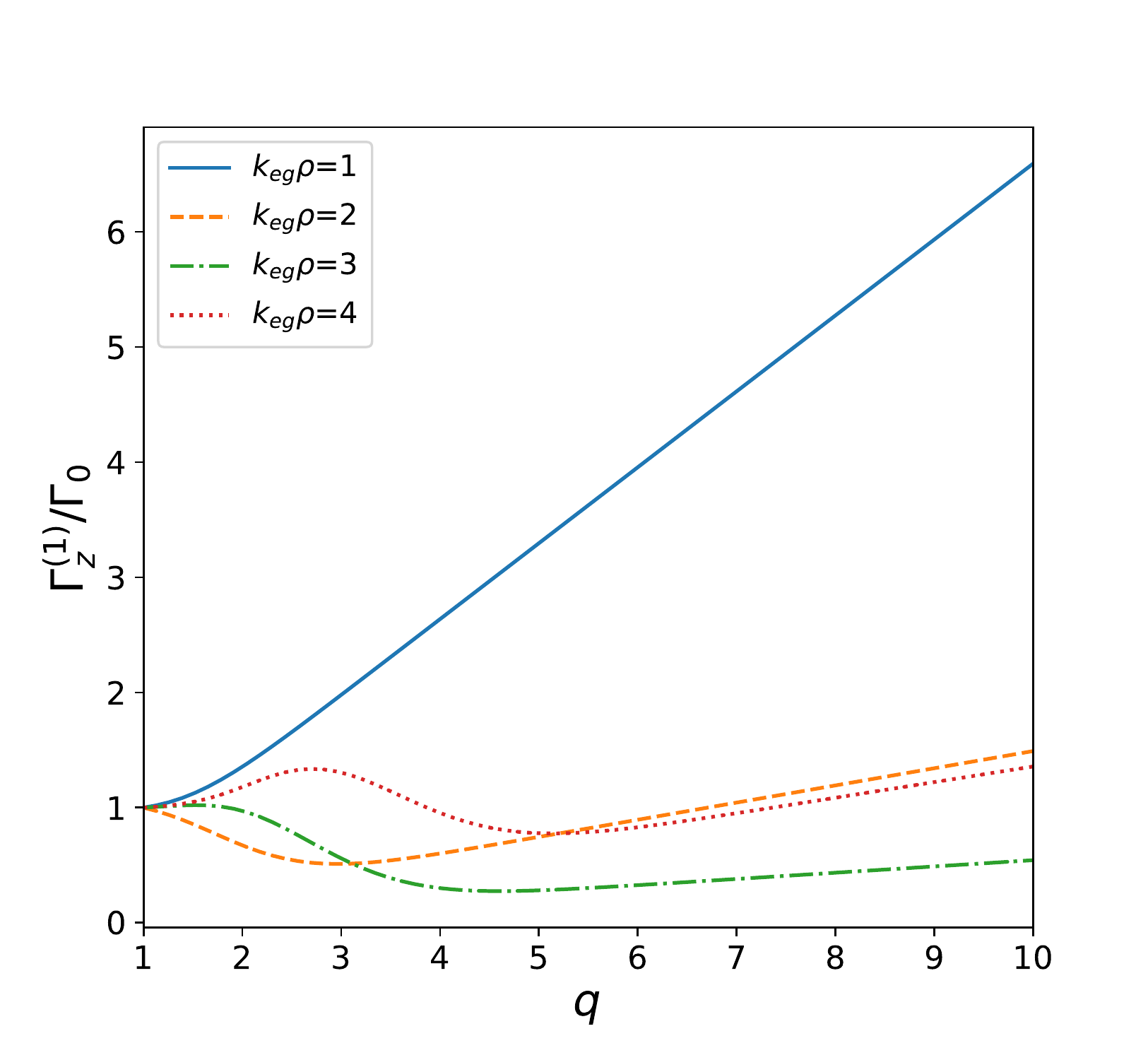}
        \caption{}
    \label{fig:Gammaz_q}
    \end{subfigure}
     \begin{subfigure}[b]{\columnwidth}
        \centering
        \includegraphics[width=\textwidth]{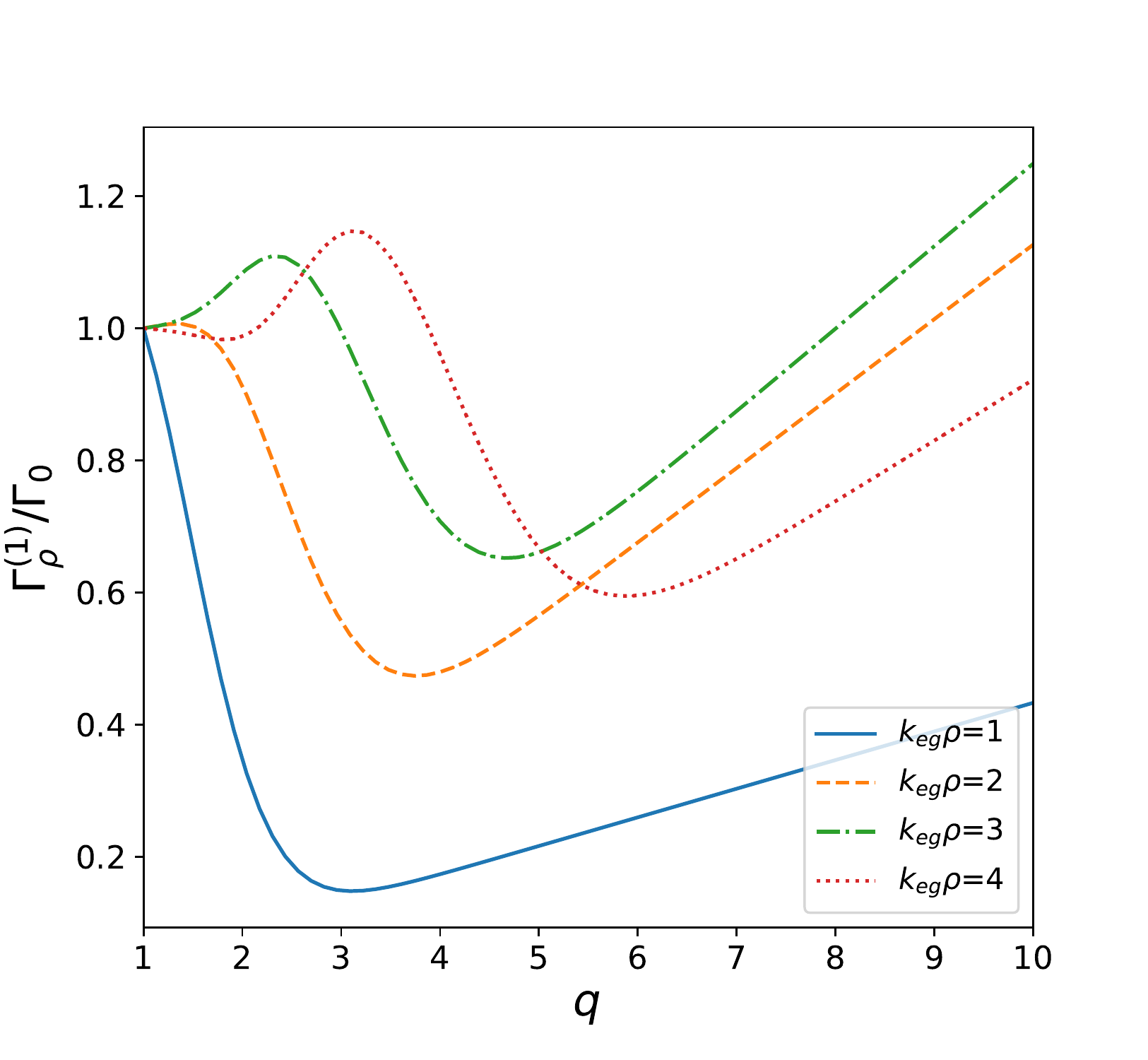}
        \caption{}
        \label{fig:Gammarho_q}
     \end{subfigure}
     \\
     \begin{subfigure}[b]{\columnwidth}
        \centering
        \includegraphics[width=\textwidth]{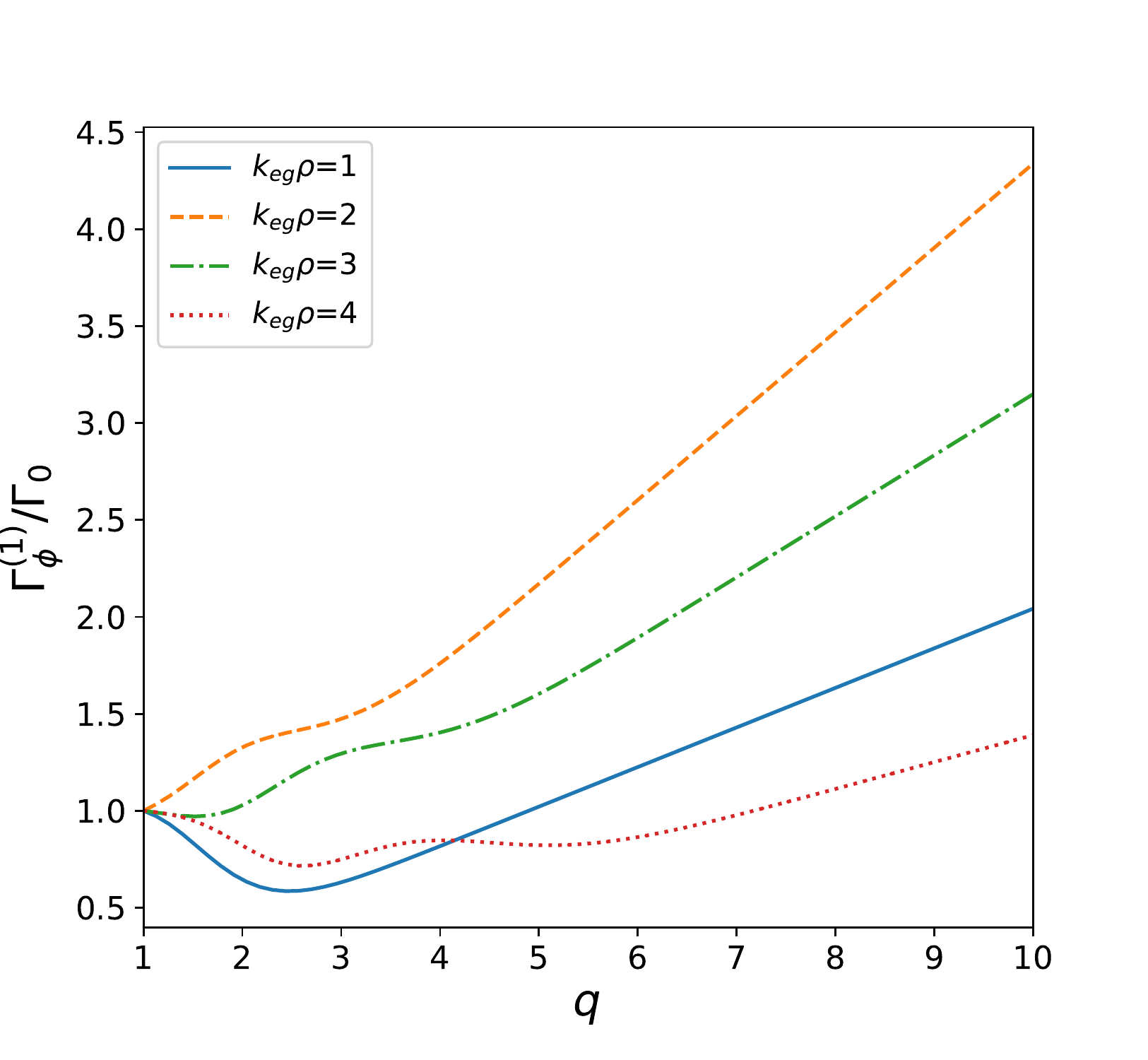}
        \caption{}
        \label{fig:Gammaphi_q}
     \end{subfigure}
     \begin{subfigure}[b]{\columnwidth}
        \centering
        \includegraphics[width=\textwidth]{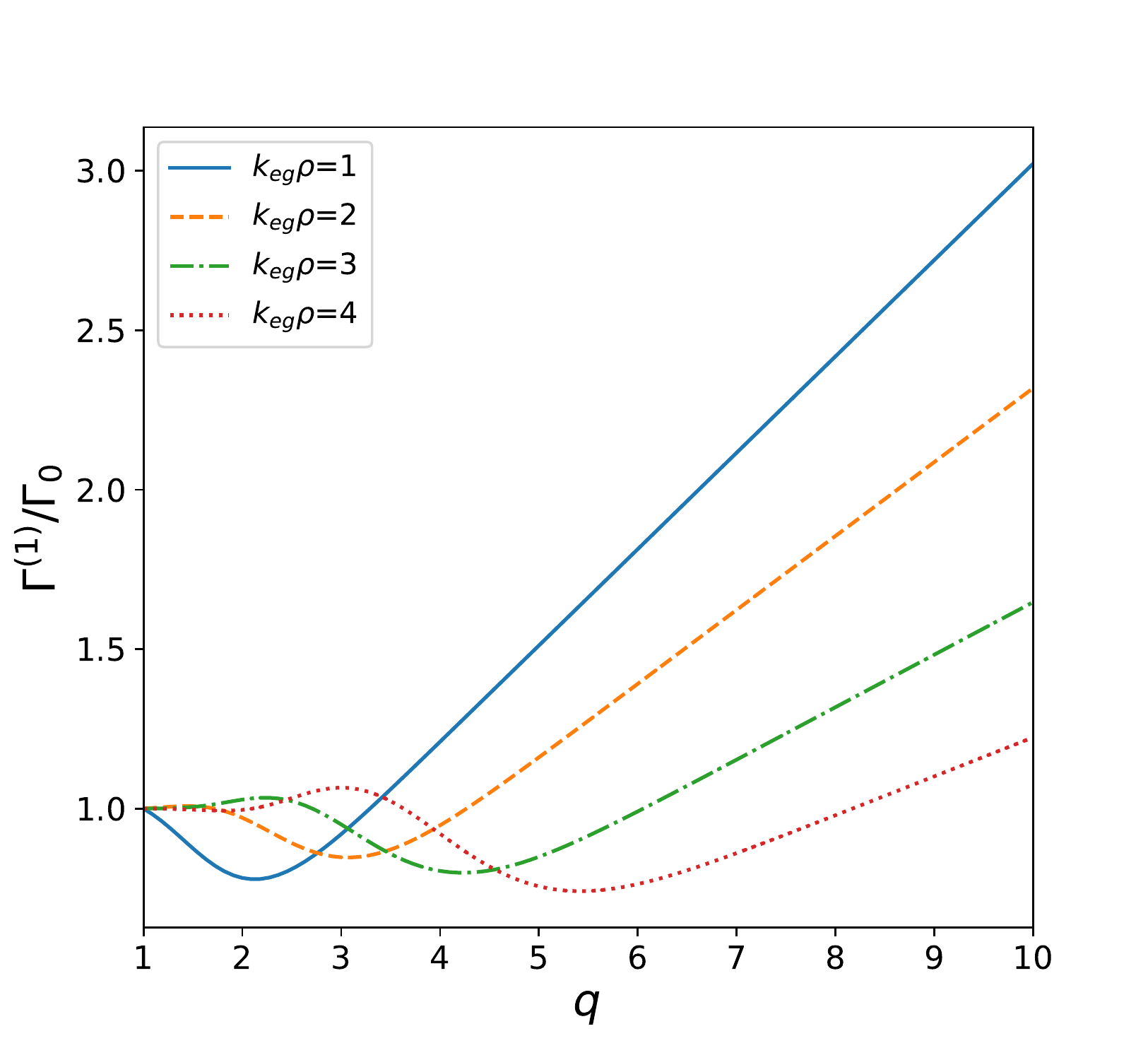}
        \caption{}
        \label{fig:Gammaiso_q}
    \end{subfigure}
    \caption{OPSE rates, as a function of $q$, of the quantum emitter, when its dipole moment is: (\subref{fig:Gammaz_r}) aligned with the direction of the string; (\subref{fig:Gammarho_r}) aligned with the radial direction and perpendicular to the string; (\subref{fig:Gammaphi_r}) is aligned with the tangential direction and perpendicular to the string; (\subref{fig:Gammaiso_r}) randomly oriented.}
    \label{fig:Gamma_q}
\end{figure*}


Note also that, as we increase the distance between the atom and the string, all panels of 
Fig. \ref{fig:Gamma_r} exhibit oscillations around the free-space values. This can be understood as an interference phenomenon if we decompose each  electromagnetic field mode as ${\bf A}_{{\bf k}p} = {\bf A}_{{\bf k}p}^{(0)} + {\bf A}_{{\bf k}p}^{(sca)}$, where ${\bf A}_{{\bf k}p}^{(0)}$ is the corresponding free-space  mode and  all the influence of the cosmic string for this field mode is encoded in the second term, ${\bf A}_{{\bf k}p}^{(sca)}$. 
 Since the field modes must be evaluated at the emitter's position (see Eq. (3)), we see that  depending on the distance between the emitter and the cosmic string the free-space contribution (${\bf A}_{{\bf k}p}^{(0)}$) and the cosmic string dependent term (${\bf A}_{{\bf k}p}^{(sca)}$)  may interfere constructively or destructively.

Finally, note that when the distance between the atom and the string goes to zero the only non-zero contribution for the SE rate comes from the orientation in which the transition dipole moment points in the $\vu z$ direction (parallel to the string). This fact may be understood as follows: 
 in this limit, the field modes have their electric fields parallel to the string so that the only non-zero coupling between the electric field and the transition dipole moment occurs with such orientation. The previous statement can be checked by taking the appropriate limit in Eqs. \eqref{eq:Amodo0} and \eqref{eq:Amodo1} and noting that the components of the electric field perpendicular  to the string involve only derivatives of the Bessel functions which goes to zero at the origin since they can be written as a linear combination of two Bessel functions with order $q|m|\pm 1\neq 0\;\forall m$.


In order to see more in more detail what is the behavior of the normalized SE rates for short distances, we shall use well known approximations of the Bessel functions involved. For small $k_{eg}\rho$, one can approximate the Bessel functions of first kind as
\begin{align}
    J_\nu(z)&\approx \frac1{2^\nu}\frac{z^\nu}{\Gamma(\nu+1)},\;\;\;\text{if}\;\;\nu\neq0;\;\;\; \vert z\vert\ll 1; \\
    J_0(z)&\approx 1-\frac{z^{2}}4;
\end{align}
so that, in the case where the dipole is aligned with the string, we obtain
\begin{align}
    \label{gamazaprox}
    \frac{\Gamma_{\vu z}}{\Gamma_0}&\approx\frac{3q}2\int_0^1\dd u\frac{u^3}{\sqrt{1-u^2}}\left[J^2_0(k_{eg}\rho)+2J^2_1(k_{eg}\rho)\right]\nonumber \\
    &\approx\frac{3q}2\int_0^1\dd u\frac{u^3}{\sqrt{1-u^2}}\nonumber\\
    &\times\left[1-\frac12(k_{eg}\rho u)^2+\frac1{2^{2q-1}}\frac{(k_{eg}\rho u)^{2q}}{\Gamma^2(q+1)}\right]\nonumber\\
    &\approx q\left[1-\frac25(k_{eg}\rho)^2+\frac{3(q+1)(k_{eg}\rho)^{2q}}{(q+\frac32)\Gamma(2q+2)}\right]\, .
\end{align}
Recalling that $q>1$ ($q=1$ corresponds to free-space), the above result shows that, up to second order terms, the OPSE rate behaves as an inverted parabola, as can be seen by a direct inspection in Fig. \ref{fig:Gammaz_r}.

%
%
The behaviors of $\Gamma_{\vu*\phi}$ and $\Gamma_{\vu*\rho}$ for an atom located very close to the cosmic string follow in a  similar way:
\begin{align}
    \frac{{\Gamma}_{\vu*\rho}}{\Gamma_0}
    &\approx q\left[\frac1{20}(k_{eg}\rho)^2+\frac{2(q+1)}{(q+\frac12)\Gamma(2q)}(k_{eg}\rho)^{2(q-1)}\right],\label{gamarhoaprox}\\
    \frac{{\Gamma}_{\vu*\phi}}{\Gamma_0}
    &\approx q\left[\frac14(k_{eg}\rho)^2+\frac{2(q+1)}{(q+\frac12)\Gamma(2q)}(k_{eg}\rho)^{2(q-1)}\right],\label{gamaphiaprox}
\end{align}
where we used that $J_{n}(x)=(-1)^{n}J_{-n}(x)$ for $n\in \mathbb{Z}$. 

The plots in Figs. \ref{fig:Gammaz_q}, \ref{fig:Gammarho_q} and \ref{fig:Gammaphi_q} depict the same OPSE rates but now as  functions of $q$ for different values of $k_{eg}\rho$. It is noticeable that, for a fixed position of the quantum emitter,  the OPSE rates do not exhibit a monotonic behavior  with respect to  parameter $q$. In fact, this quite subtle  behavior  can be explained qualitatively by an argument analogous to that one used to explain the oscillations of the OPSE rates with the distance between the emitter and the string, but this time  we shall look for an interference  pattern along the angular coordinate. Again, we can write the field mode as a sum of two contributions. The first one is the free-space contribution and the second one encodes the  influence of the string. At this moment it is convenient to remember  that the presence of the string gives rise to a flat spacetime with a conical singularity, which is equivalent to say that there is a deficit angle (denoted by $\delta\phi$). For a fixed $\rho$, 
whenever $(2\pi - \delta\phi)\rho = n \lambda_{eg}$, with $n$ a positive integer, the two contributions will interfere constructively, enhancing the OPSE rate. By writing $\delta\phi$ in terms of $q$ in the previous relation, we have constructive interference for every $q_n$ such that $2\pi\rho/q_n = n \lambda_{eg}$.  As $q$ is increased from $q_n$,  this kind of interference condition is not satisfied anymore and the OPSE rate diminishes until $q = q_{n-1}$, where we have another enhancement of the OPSE rate. Hence, as $q$ is continuously increased we expect some oscillations in the OPSE rate. However,  since  $n$ can not be smaller than $1$, no interference pattern is observed in the region $q > q_1 = k_{eg}\rho$, only a monotonic behavior of the OPSE rate as a function of $q$. We also emphasize that since $q$ must be greater than $1$, we have a finite number of oscillations 

It is worth mentioning that, whenever the condition $q\gtrsim k_{eg}\rho$ is satisfied, a linear dependence of the OPSE rates with parameter $q$ (for any orientation of the transition dipole moment) shows up for any distance between the cosmic string and the quantum emitter. 
This can be seen mathematically with the aid of  the following approximation for  Bessel functions \cite{abramowitz},
\begin{equation}
    J_\nu(z)\approx\frac{e^\nu}{\sqrt{2\pi \nu}}\left(\frac{z}{2\nu}\right)^\nu,\;\;\; \nu\gg 1\;\;\; \text{if}\;\;z\neq 0. 
\end{equation}
Notice that, in the base of the power $\nu$, when $\nu\gg z$, the Bessel function rapidly decreases as $\nu$ increases. As a consequence, in the summations written in Eqs. \eqref{eq:Gammaz}, \eqref{eq:Gammarho} and \eqref{eq:Gammaphi}, the terms with $|m|\neq0$ may be neglected if $q\gtrsim k_{eg}\rho$, since for a fixed $k_{eg}\rho$ there is a power decay when $q$ increases. Thus we are left with
\begin{align}
    \frac{{\Gamma}_{\vu*z}}{\Gamma_0}&\approx\frac{3q}2\int_0^1\frac{\dd u\;u^3}{\sqrt{1-u^2}}J^2_0(k_{eg}\rho u)\propto q,\label{aprox1}\\
    \frac{{\Gamma}_{\vu*\rho}}{\Gamma_0}&\approx\frac{3q}2\int_0^1\frac{\dd u\;u}{\sqrt{1-u^2}} J^2_1(k_{eg}\rho u)\propto q,\label{aprox2}\\
    \frac{{\Gamma}_{\vu*\phi}}{\Gamma_0}&\approx\frac{3q}2\int_0^1\frac{\dd u\;u}{\sqrt{1-u^2}}J^2_1(k_{eg}\rho u)\propto q.\label{aprox3}
\end{align}
All the previous statements are in qualitative agreement with the plots shown in Figs. \ref{fig:Gammaz_q}, \ref{fig:Gammarho_q} and \ref{fig:Gammaphi_q}.
%


\section{\label{sec:twophotonSS}Two-Photon Spontaneous Emission}

In this section we consider the same physical system as that shown in Fig. \ref{fig:setup}, which consists of an atom, initially in one of its excited states, near a cosmic string in the free-space, but now we analyze the TPSE instead of the OPSE.  In contrast to the latter case, TPSE is characterized by a broadband spectrum of emission, where the frequencies of the two emitted photons, say $\omega_1$ and $\omega_2$, are allowed to have any continuous value satisfying the energy conservation condition, namely, $\omega_1 + \omega_2 = \omega_{eg}$. Due to this feature, we will  be interested not only in the total decay rate but also in the probability density function for a photon to be emitted with a frequency within the interval $[\omega,\omega+\dd\omega]$. As expected from the Purcell effect for OPSE rate, both quantities are modified by the presence of the string and provide signatures that may help in the attempts to observe cosmic strings.

In the following we start by describing the general theory behind the TPSE, which comes directly from second-order perturbation theory.  Then, after some simplifications, we show that the OPSE rates previously calculated play an essential role that allows us to calculate the TPSE rate. Subsequently, we  apply this formalism in the system under consideration, i.e. an atom in the vicinities of the cosmic string.

\subsection{Methodology}

As in the case of OPSE we consider the electric dipole approximation, with $H_{\text{int}}$ written as in Eq. \eqref{eq:Hint}, but now, since we are interested in calculating TPSE rates,  it is necessary to employ a second-order perturbation theory. The reader is referred to Ref.  \cite{muniz2019} for a more detailed discussion of this topic. 

As before the atom is  considered  initially in an excited state and there are no photons in the field. Such an initial state is denoted by $|i\rangle=|e;0\rangle$. After the atomic transition, the atom is in a lower energy level, and there are  two photons in the field, in the modes $\alpha$ and $\alpha'$. We represent this final state by $\ket{g; 1_{\alpha},1_{\alpha'}}$. Since we shall employ a second-order perturbation theory, besides the previous states we must consider all intermediate states $\ket I$ that are connected to the initial and final states through the interaction Hamiltonian $H_{\text{int}}$, namely,  all  states whose transition matrix elements with the initial and final states are different from zero. It is not difficult to see that the states $\ket I$ must contain one photon in the field so that we denote them by $\ket{m; 1_{\alpha}}$ or $\ket{m; 1_{\alpha'}}$, where $m$ is an index that represents the intermediate state of the atom. Thus, summing over the $m$ states and the final states of the field, one finds that
\begin{align}
    \label{eq:tpse}
    \Gamma(\vb{r})&=\frac{\pi}{4\epsilon_{0}^{2}\hbar^{2}}\sum_{\alpha, \alpha'}\omega_{\alpha}\omega_{\alpha'}|\vb{A}_{\alpha}({\bf r})\cdot\mathbb{D}(\omega_{\alpha},\omega_{\alpha'})\cdot \vb{A}_{\alpha'}({\bf r})|^{2}\nonumber \\
    &\times\delta(\omega_{\alpha}+\omega_{\alpha'}-\omega_{eg}),
\end{align}
%
where we have defined the tensor
\begin{equation}
    \mathbb D (\omega_{\alpha}, \omega_{\alpha'})\equiv\sum_{m} \left[\frac{\vb d_{em}\vb d_{mg}}{\omega_{em}-\omega_{\alpha}}+\frac{\vb d_{mg}\vb d_{em}}{\omega_{em}-\omega_{\alpha'}}\right].
\end{equation}

The subsequent discussion may be simplified considerably if we write the $\vb A_\alpha$ modes in terms of the Green dyadics which is the solution of the equation
\begin{align}
    \curl\curl\mathbb G(\vb r,\vb r';\omega)-\frac{\omega^2}{c^2}\mathbb G(\vb r,\vb r';\omega)=\mathbb I\delta(\vb r-\vb r').
\end{align}
subjected to the appropriate boundary conditions. 
More specifically, we will make use of its imaginary part, that admits the following spectral representation
\begin{equation}\label{eq:ImG}
    \Im\mathbb G(\vb r,\vb r'; \omega)=\frac{\pi c^2}{2\omega}\sum_{\alpha}\vb A_{\alpha}(\vb r)\vb A^*_{\alpha}(\vb r')\delta(\omega-\omega_\alpha).
\end{equation}
Therefore, substituting Eq. \eqref{eq:ImG} in Eq. \eqref{eq:tpse}, one finds that
it is possible to write the total spontaneous emission rate as an integral of another function,
\begin{equation}
    \Gamma(\vb r)=\int^{\omega_{eg}}_0\dd\omega\;\gamma(\vb r;\omega),
\end{equation}
where
\begin{align}\label{eq:stringTPSE}
    \gamma(\vb r;\omega)=&\frac{\mu^2_0}{\pi\hbar^2}\omega^2(\omega_{eg}-\omega)^2\Im\mathbb G_{il}(\vb r;\omega)\Im\mathbb G_{jk}(\vb r;\omega_{eg}-\omega)\nonumber\\
    &\times\mathbb D_{ij}(\omega,\omega_{eg}-\omega)\mathbb D^*_{lk}(\omega,\omega_{eg}-\omega)
\end{align}
is referred as the spectral density. Note that the spectral density  must be symmetric with respect to half of the transition frequency, $\gamma(\omega) = \gamma(\omega_{eg} - \omega)$. This is a direct consequence of energy conservation, since every time a photon is emitted with frequency $\omega$ another photon  is simultaneously emitted with frequency $\omega_{eg} - \omega$.


The previous expression is quite general and can be applied, in principle, to calculate the TPSE rate of an excited quantum emitter near an arbitrary material body. However, in some still quite general situations this equation acquires a simpler form. This occurs, for instance, whenever there is a basis that diagonalizes the Green tensor at coincident points, which happens to be the case under consideration in this work, as shown in \cite{Saharian:2011sx}. For this case,  Eq. \eqref{eq:stringTPSE} takes the form
\begin{align}\label{eq:stringTPSE_diag}
    \gamma({\bf r};\omega)=&\frac{\mu^2_0}{\pi\hbar^2}\omega^2(\omega_{eg}-\omega)^2 |\mathbb D_{ij}(\omega,\omega_{eg}-\omega)|^2\cr& 
    \times\Im\mathbb G_{ii}({\bf r},{\bf r};\omega)\Im\mathbb G_{jj}({\bf r},{\bf r};\omega_{eg}-\omega).
\end{align}
Analogously to the OPSE case, it is convenient to write an expression for the spectral density $\gamma(\omega)$ normalized by its expression in free-space, first obtained by given by M. G\"oppert-Meyer in 1931 \cite{Mayer} and given by
\begin{equation}\label{eq:minkowskiTPSE}
	\gamma_0(\omega)=\frac{\mu^2_0}{36\pi^3\hbar^2c^2}\omega^3(\omega_{eg}-\omega)^3|\mathbb D(\omega,\omega_{eg}-\omega)|^2,
\end{equation}
where we defined
\begin{align}
        |\mathbb D(\omega,\omega')|^2=\mathbb D_{ij}(\omega,\omega')\mathbb D^*_{ij}(\omega,\omega').
\end{align}
Therefore, the normalized spectral density can be cast into the form
\begin{equation}
    \label{eq:spectral}
    \frac{\gamma(\vb r; \omega)}{\gamma_0(\omega)} = \sum_{i,j=1}^3\frac{|\mathbb D_{ij}(\omega,\omega_{eg}-\omega)|^2}{|\mathbb D(\omega,\omega_{eg}-\omega)|^2}P_i(\vb r; \omega)P_j(\vb r; \omega_{eg}-\omega)\, .
\end{equation}
The previous expression is also referred to as spectral enhancement \cite{rivera2017making}. The $P_i$ functions are the Purcell factors, defined by
\begin{equation}
    \label{eq:purcell}
    P_i(\vb r;\omega)\equiv\frac{6\pi c}{\omega}\Im\mathbb G_{ii}(\vb r,\vb r;\omega).
\end{equation}
Recalling Eq. \eqref{eq:ImG} for the explicit expression of the imaginary part of the Green dyadics and comparing it with Fermi's golden rule in Eq. \eqref{eq:Gamma2}, we see  that these factors are precisely the normalized OPSE rates presented in Eqs. \eqref{eq:Gammaz}, \eqref{eq:Gammarho} and \eqref{eq:Gammaphi} but now with a dependence in the frequency, as discussed in Ref. \cite{Novotny}. 

Finally, we  assume that the TPSE is originated from a $s\rightarrow s$ atomic transition. If this is the case, it can be shown that
%
%
\begin{align}
    \frac{|\mathbb D_{ij}(\omega,\omega_{eg}-\omega)|^2}{|\mathbb D(\omega,\omega_{eg}-\omega)|^2}=\frac13\delta_{ij},
\end{align}
and consequently the expression for the spectral enhancement is greatly simplified, taking the form
\begin{equation}
    \frac{\gamma(\vb r; \omega)}{\gamma_0(\omega)}=\frac13\sum_{i}P_i(\vb r; \omega)P_i(\vb r; \omega_{eg}-\omega).\label{eq:dois_fotons_final}
\end{equation}
As we shall see in the next subsection, the previous equation  allows us to compute the TPSE rate of an emitter close to a cosmic string using the Purcell factors given by Eqs. \eqref{eq:Gammaz}, \eqref{eq:Gammarho} and \eqref{eq:Gammaphi}.


\subsection{TPSE in the background of a cosmic string: results and discussions}

\begin{figure}
     \centering
     \begin{subfigure}[b]{\columnwidth}
        \centering
        \includegraphics[width=\textwidth]{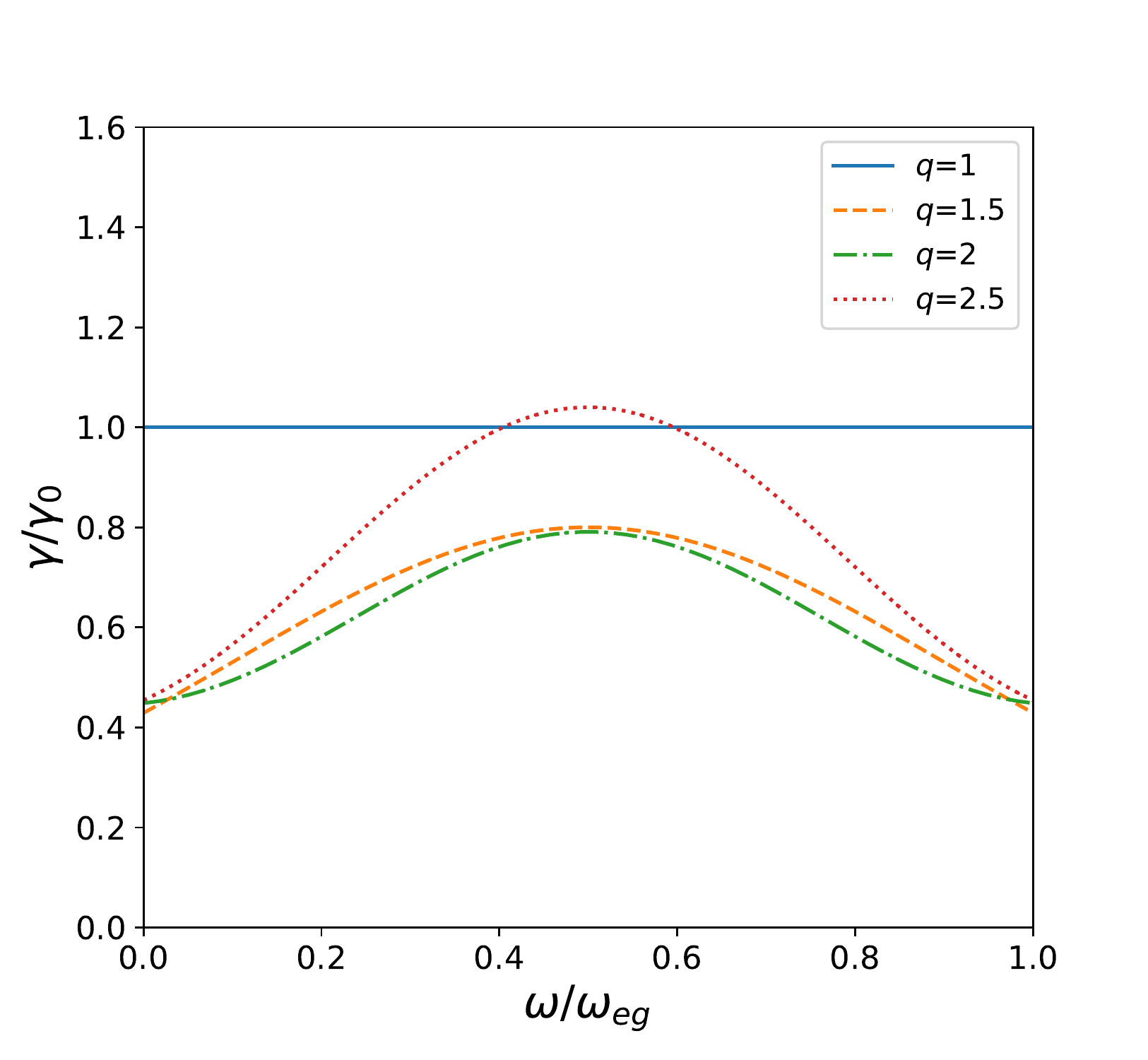}
        \caption{}
        \label{fig:k-2-por-omega}
    \end{subfigure}
    \\
    \begin{subfigure}[b]{\columnwidth}
        \centering
        \includegraphics[width=\textwidth]{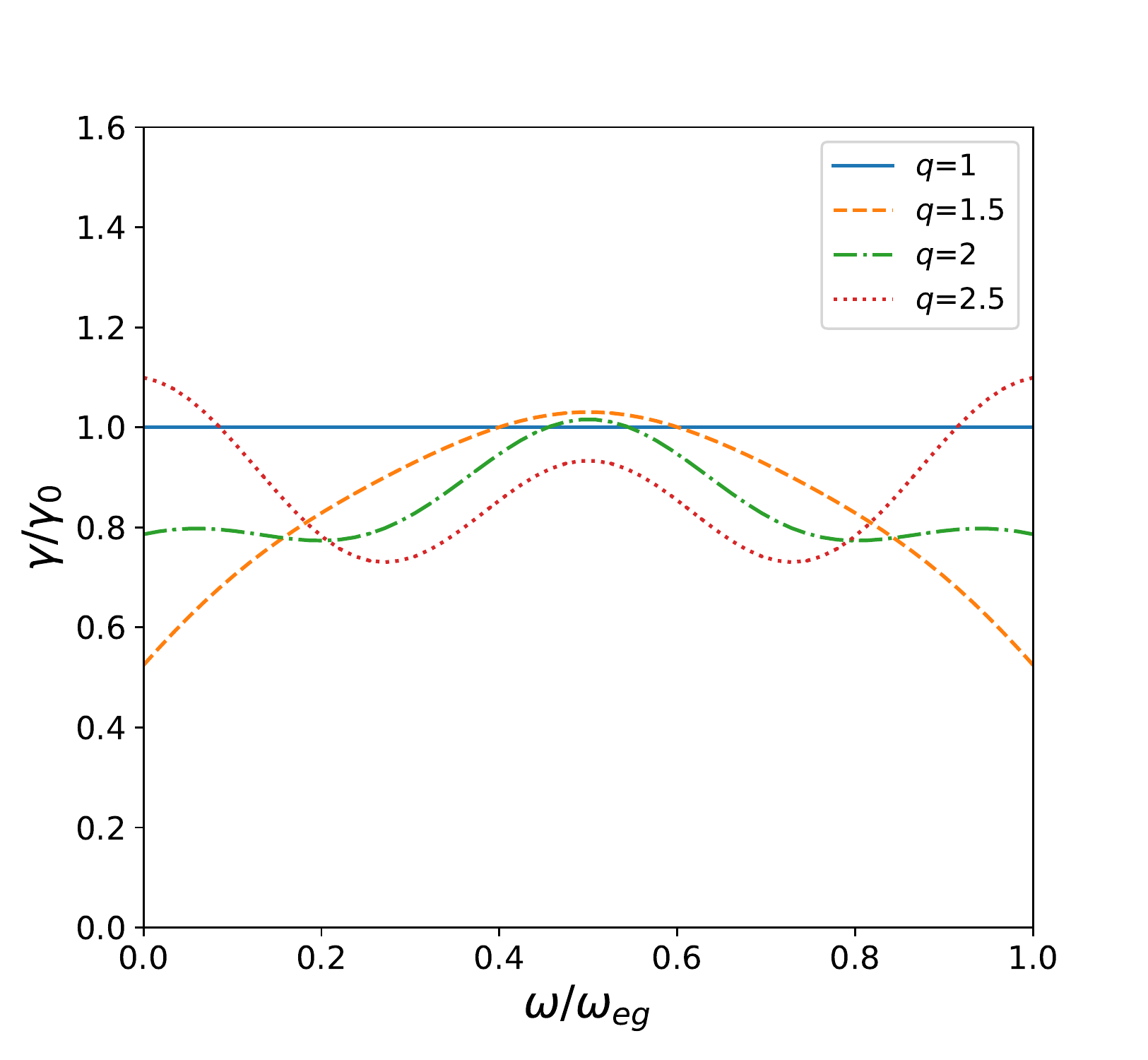}
        \caption{}
        \label{fig:k-4-por-omega}
    \end{subfigure}
    \caption{Spectral enhancement $\gamma(\omega;\vb{r})/\gamma_{0}(\omega)$ for an atom near a cosmic string as a function of the normalized frequency, for different values of $q$, setting: (\subref{fig:k-2-por-omega}) $k_{eg}\rho=2$, (\subref{fig:k-4-por-omega}) $k_{eg}\rho=4$.}
    \label{fig:doisfotons_omega_q}
\end{figure}

We are now able to investigate the TPSE rate of an atom in the vicinities of a cosmic string by calculating its spectral enhancement, by means Eq. \ref{eq:dois_fotons_final}. 
In Fig. \ref{fig:k-2-por-omega} we plot the spectral enhancement as a function of the 
normalized frequency $\omega/\omega_{eg}$ for a fixed distance between the emitter and the string ($k_{eg}\rho = 2$) for different values of $q$. Recalling that $q=1$ corresponds free-space to the free-space limit, we have  $\gamma/\gamma_0 = 1$ for this case,  which was included in  Fig. \ref{fig:k-2-por-omega}  just for comparison. Looking for the other curves, with  $q\ne 1$, we see that the presence of the string indeed alters the spectral density of the emitted photons. Note that, as expected, all curves in this figure are symmetrical with respect to $\omega/\omega_{eg} = 1/2$. Depending on the value of the emitted frequency and the value of $q$, we may have an enhancement or an attenuation of the spectral density. As we shall see, the spectral enhancement at a given frequency has a non-monotonic behavior as we increase the density mass of the string (or, equivalently, as we increase $q$). 
 Fig. \ref{fig:k-4-por-omega} depicts a completely  analogous situation as that shown in Fig. \ref{fig:k-2-por-omega}  but now with $k_{eg}\rho=4$. 
 Note that, in average, the curves in  Fig. \ref{fig:k-4-por-omega}  are closer to the unity value (corresponding to the free-space situation) than the curves in Fig. \ref{fig:k-2-por-omega}. This is due to the fact that in  Fig. \ref{fig:k-4-por-omega} the distance from the emitter to the string is greater than in 
 Fig. \ref{fig:k-2-por-omega}. However, as we shall see later, the spectral density at a given frequency tends to the free-space value as the distance to the string increases in a non-monotonic way. Finally, the fact that $\gamma/\gamma_0$ remains finite and non zero as $\omega/\omega_{eg}\rightarrow 0$ (or equivalently $\omega/\omega_{eg}\rightarrow 1$) means that both $\gamma$ and $\gamma_0$ have the same power law with respect to $\omega$ in these limits.


\begin{figure}[h!]
     \centering
     \begin{subfigure}[b]{\columnwidth}
         \centering
        \includegraphics[width=\textwidth]{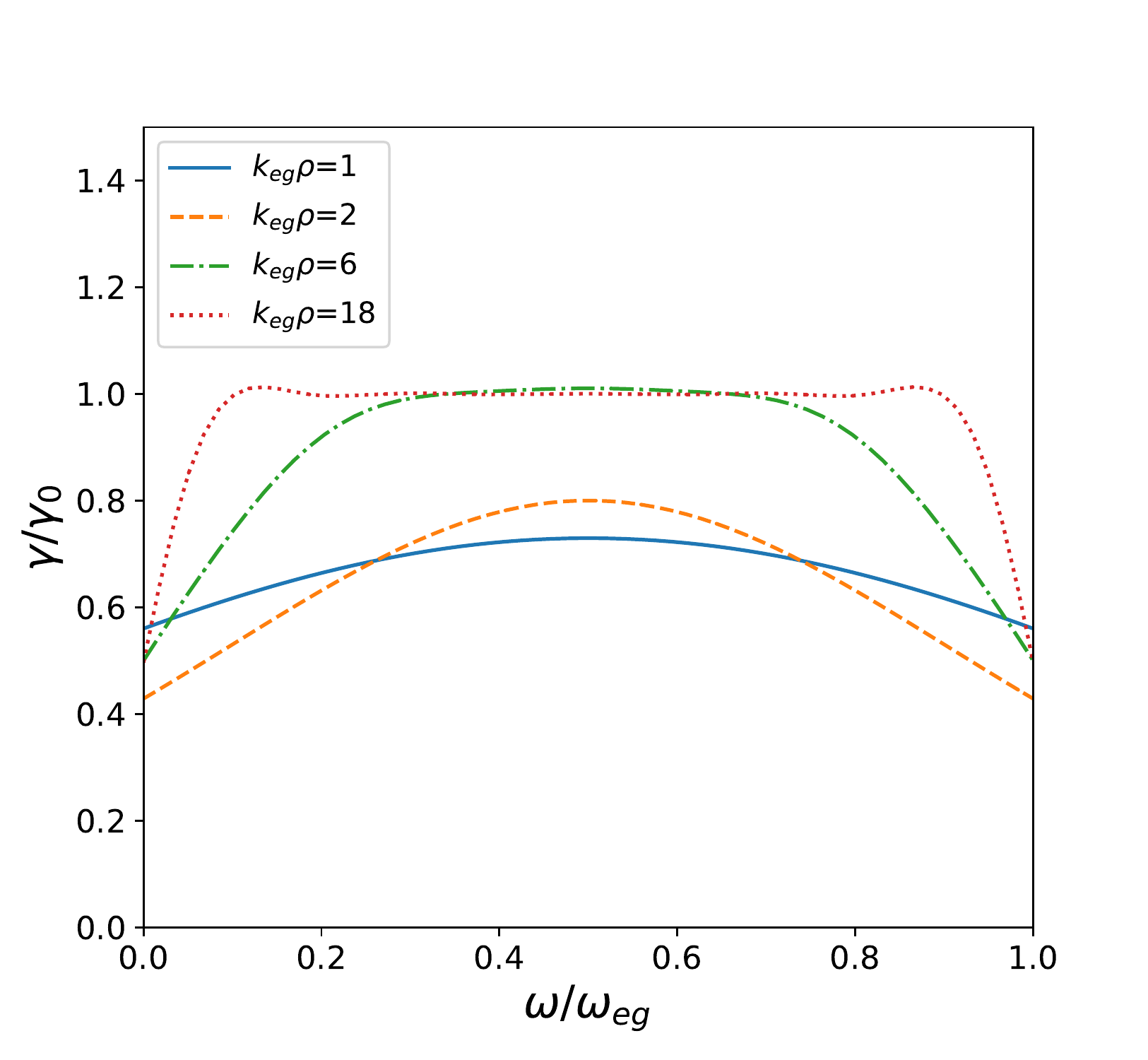}
        \caption{}
        \label{fig:q-1-5-por-omega}
    \end{subfigure}
    \\
    \begin{subfigure}[b]{\columnwidth}
        \centering
        \includegraphics[width=\textwidth]{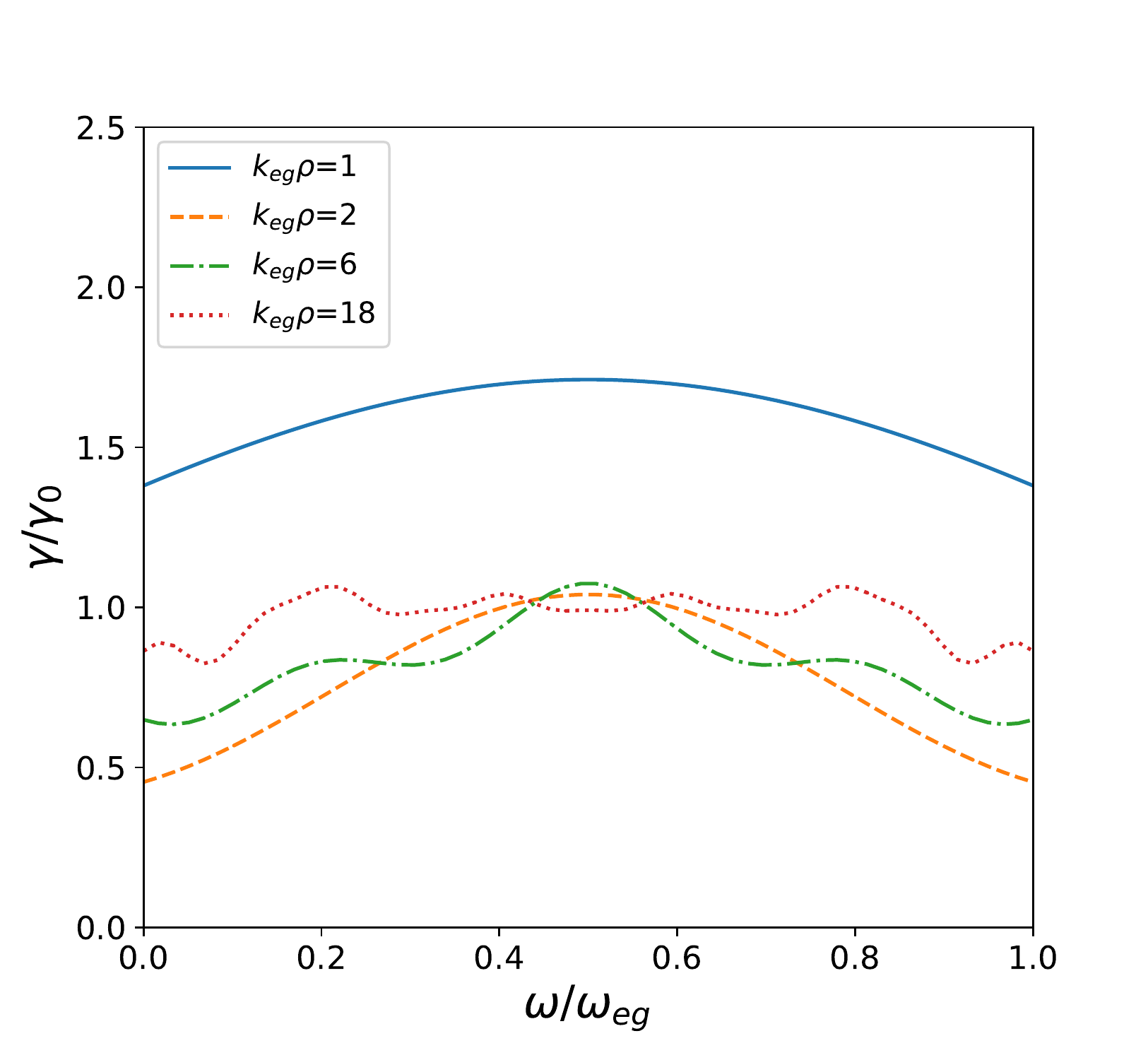}
        \caption{}
        \label{fig:q-2-5-por-omega}
    \end{subfigure}
    \caption{Spectral enhancement $\gamma(\omega;\vb{r})/\gamma_{0}(\omega)$ for an atom near a cosmic string as a function of the normalized frequency, for different values of $k_{eg}\rho$, setting: (\subref{fig:q-1-5-por-omega}) $q=1.5$, (\subref{fig:q-2-5-por-omega}) $q=2.5$.}
    \label{fig:doisfotons_omega_k}
\end{figure}

\begin{figure}
     \centering
     \begin{subfigure}[b]{\columnwidth}
         \centering
        \includegraphics[width=\textwidth]{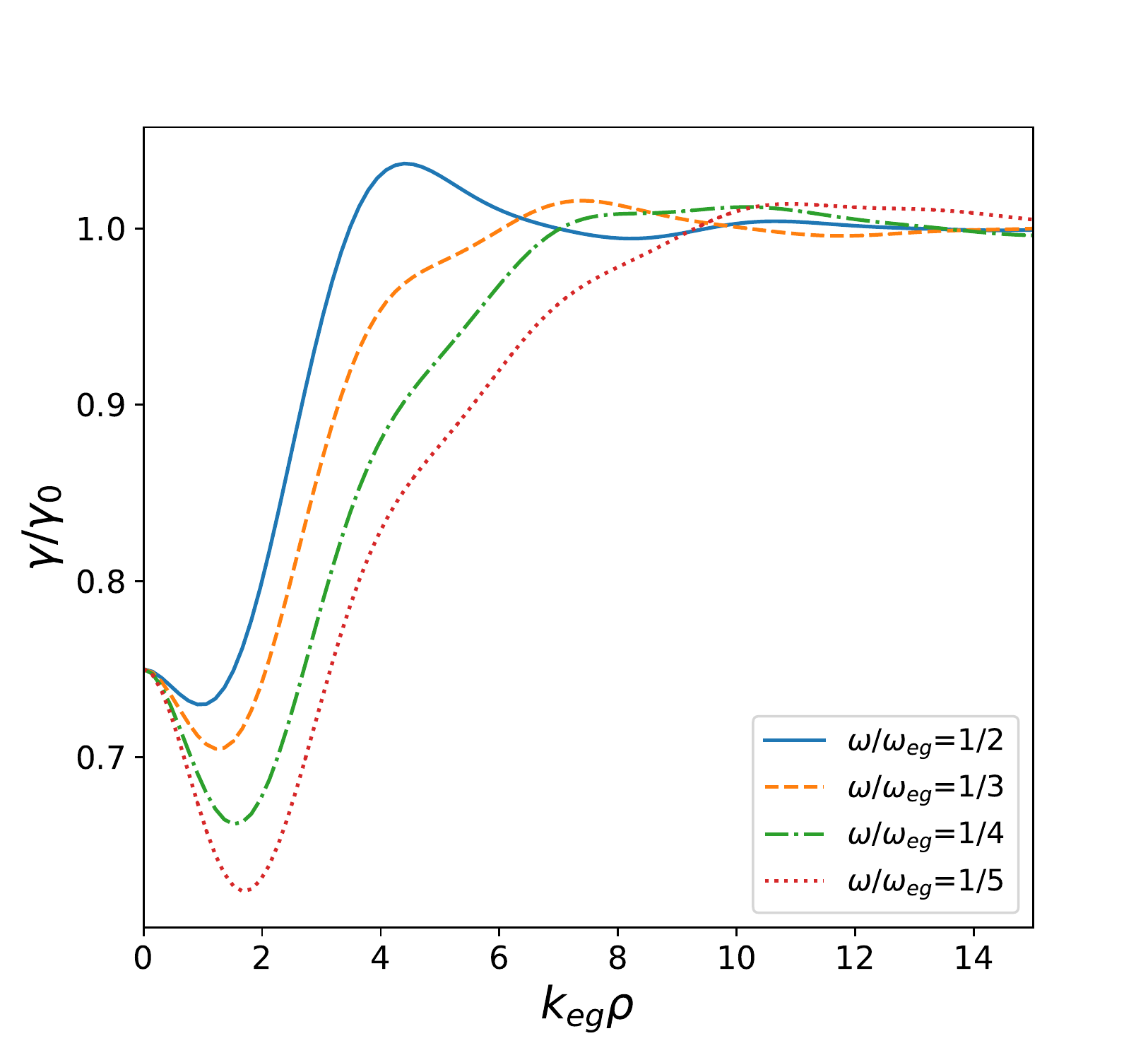}
        \caption{}
        \label{fig:q-1-5-por-distancia}
    \end{subfigure}
    \\
    \begin{subfigure}[b]{\columnwidth}
        \centering
        \includegraphics[width=\textwidth]{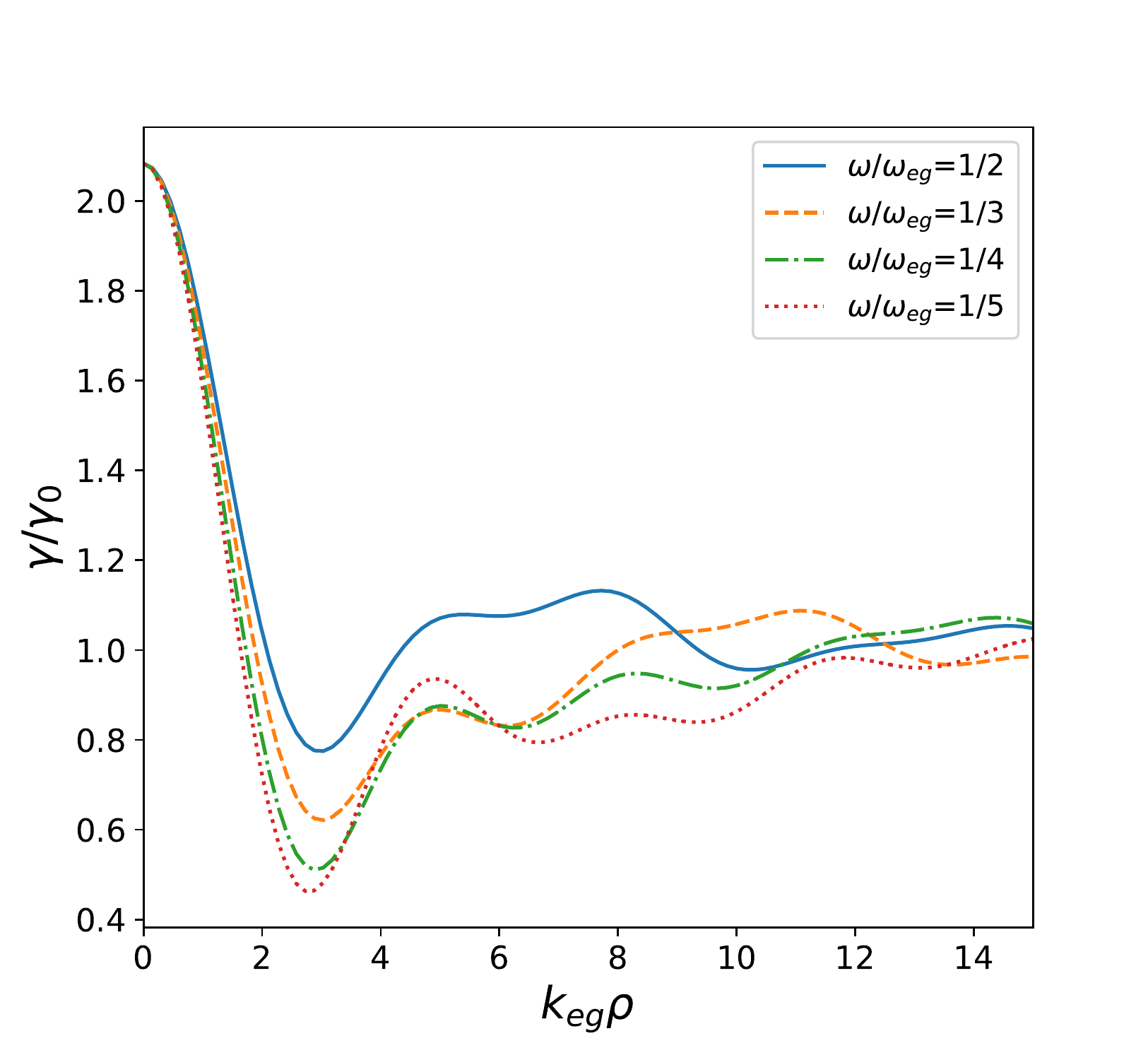}
        \caption{}
        \label{fig:q-2-5-por-distancia}
     \end{subfigure}
    \caption{Spectral enhancement $\gamma(\omega;\vb{r})/\gamma_{0}(\omega)$ for an atom near a cosmic string as a function of the normalized distance, for different values of $\omega/\omega_{eg}$, setting: (\subref{fig:q-1-5-por-distancia}) $q=1.5$, (\subref{fig:q-2-5-por-distancia}) $q=2.5$.}
    \label{doisfotons_distancia}
\end{figure}

In Fig. \ref{fig:q-1-5-por-omega} we plot the spectral enhancement as a function of the normalized frequency $\omega/\omega_{eg}$ for $q=1.5$ and different values of the normalized distance $k_{eg}\rho$ between the emitter and the string. Note also that as  $k_{eg}\rho$ increases, the presence of the string becomes less important, 
so that $\gamma/\gamma_0\to1$ far from the string. This is quite evident in this figure  for the values 
$k_{eg}\rho=6$  and $k_{eg}\rho=18$. However, as already mentioned, there will be a non-monotonic behavior in way the spectral density at a given frequency tends to its free-space value as we increase $k_{eg}\rho$, as it will become evident in Figs. \ref{fig:q-1-5-por-distancia}  and \ref{fig:q-2-5-por-distancia}. 

\begin{figure}
    \centering
    \includegraphics[width=\columnwidth]{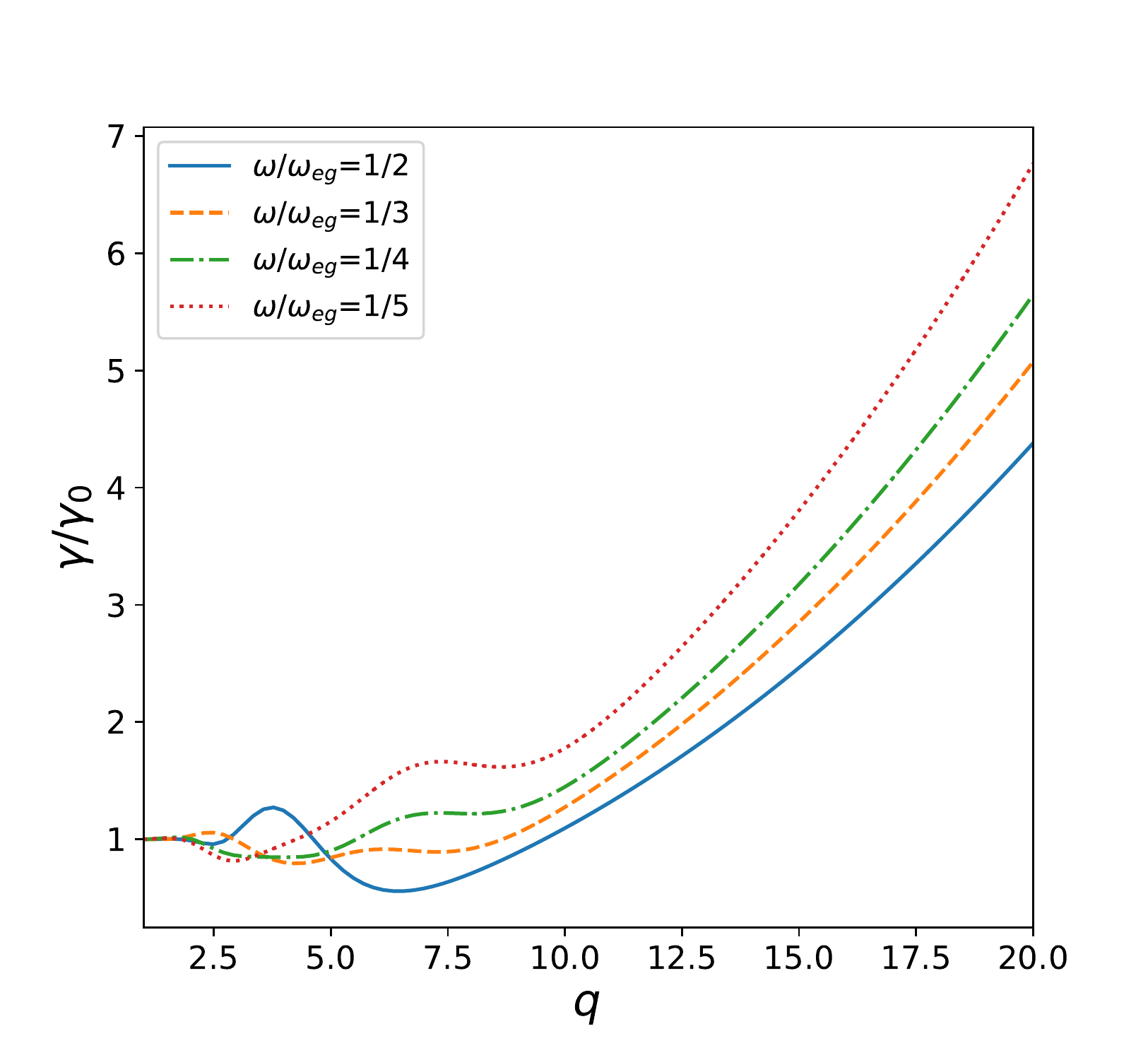}
    \caption{Spectral enhancement $\gamma(\omega;\vb{r})/\gamma_{0}(\omega)$ of an atom near a cosmic string as a function of $q$ for different frequencies, setting $k_{eg}\rho=10$.}
    \label{fig:k-10-por-q}
\end{figure}

In Fig. \ref{fig:q-2-5-por-omega} we plot the same curves as in Fig. \ref{fig:q-1-5-por-omega},  but with a different value of $q$. While in the latter  we chose $q=1.5$, in the former we used $q=2.5$. Since increasing $q$ means to increase the deficit angle and, in principle, to increase the influence of the string, we see  that, as $k_{eg}\rho$ increases, the spectral enhancement $\gamma/\gamma_0$ tends to $1$ slower in Fig. \ref{fig:q-2-5-por-omega} than in Fig. \ref{fig:q-1-5-por-omega}. Nevertheless, as we shall see in Fig. \ref{fig:k-10-por-q}, $\gamma/\gamma_0$ has a non-monotonic behavior as a function of $q$.

It is interesting to observe that, as the quantum emitter is moved away from the string, the spectral enhancement for frequencies 
around $\omega/\omega_{eg} = 1/2$  approaches  1  (free-space value) faster than  the spectral enhancement for frequencies near zero or 
$\omega/\omega_{eg} = 1$. We can understand qualitatively this behavior as follows. Photons with small frequencies (near zero, for instance) have wavelengths greater than photons with frequency 
$\omega = \omega_{eg}/2$, so that the spectral enhancement  for these small frequencies will approach their corresponding free-space values slower than the spectral enhancement for frequencies around $\omega/\omega_{eg} = 1/2$. Naively, one could think that for frequencies near the maximum value $\omega_{eg}$ the opposite would occur. However, this is not the case since for an emitted photon with frequency $\omega$, another one is simultaneously emitted with a frequency $\omega_{eg} - \omega$.

The $k_{eg}\rho$ dependence of the spectral enhancement is depicted in Figs. \ref{fig:q-1-5-por-distancia} (for $q=1.5$) and \ref{fig:q-2-5-por-distancia} (for $q=2.5$) for different (fixed) values of $\omega/\omega_{eg}$. Note that, in close analogy to what happens in the OPSE case, the spectral enhancement at a given frequency oscillates around 1 as $k_{eg}\rho$ is increased and tends to 1 in the limit $k_{eg}\rho \rightarrow \infty$, as expected. However, the oscillations present in  \ref{fig:q-1-5-por-distancia} are more irregular than those appearing in Fig. \ref{fig:Gamma_r} for the OPSE case. The reason for that is related to the fact that in the TPSE case another length scale is present since, now, two photons are emitted and hence we have two different wavelengths, except when the two photons are emitted with frequency $\omega_{eg}/2$. The above statements can be seen by a direct inspection for instance in Fig. \ref{fig:q-1-5-por-distancia}.

In order to discuss the behavior of the spectral enhancement for $k_{eg}\rho\ll 1$ in more detail, we  use 
the approximated expressions of the OPSE rates previously calculated in Eqs. \eqref{gamazaprox}, \eqref{gamarhoaprox} and \eqref{gamaphiaprox}. For $q\neq 1$ and up to order $(k_{eg}\rho)^{2}$ we have
    \begin{align}
        &\frac{\gamma}{\gamma_0}\approx\frac{q^2}3\left\{1-\frac{2(k_{eg}\rho)^2}5\left[\left(\frac{\omega}{\omega_{eg}}\right)^{2}+\left(1-\frac{\omega}{\omega_{eg}}\right)^2\right]\right\}\nonumber\\
        &+\mathcal O[(k_{eg}\rho)^{4(q-1)}]
    \end{align}
   This result implies that a parabolic behavior for small $k_{eg}\rho$ shows up, which can be seen in  Figs. \ref{fig:q-1-5-por-distancia} and \ref{fig:q-2-5-por-distancia}.
    
It is worth emphasizing another interesting features.  First, for $k_{eg}\rho\ll1$ the spectral enhancement is independent of the frequency $\omega/\omega_{eg}$, which means that for short distances the spectral density is  proportional to the free-space spectral density. This property can be seen in Figs. \ref{fig:q-1-5-por-distancia} and \ref{fig:q-2-5-por-distancia} in the region where the former limit is satisfied. 

\begin{figure}
    \centering
    \begin{subfigure}[b]{\columnwidth}
        \centering
        \includegraphics[width=\textwidth]{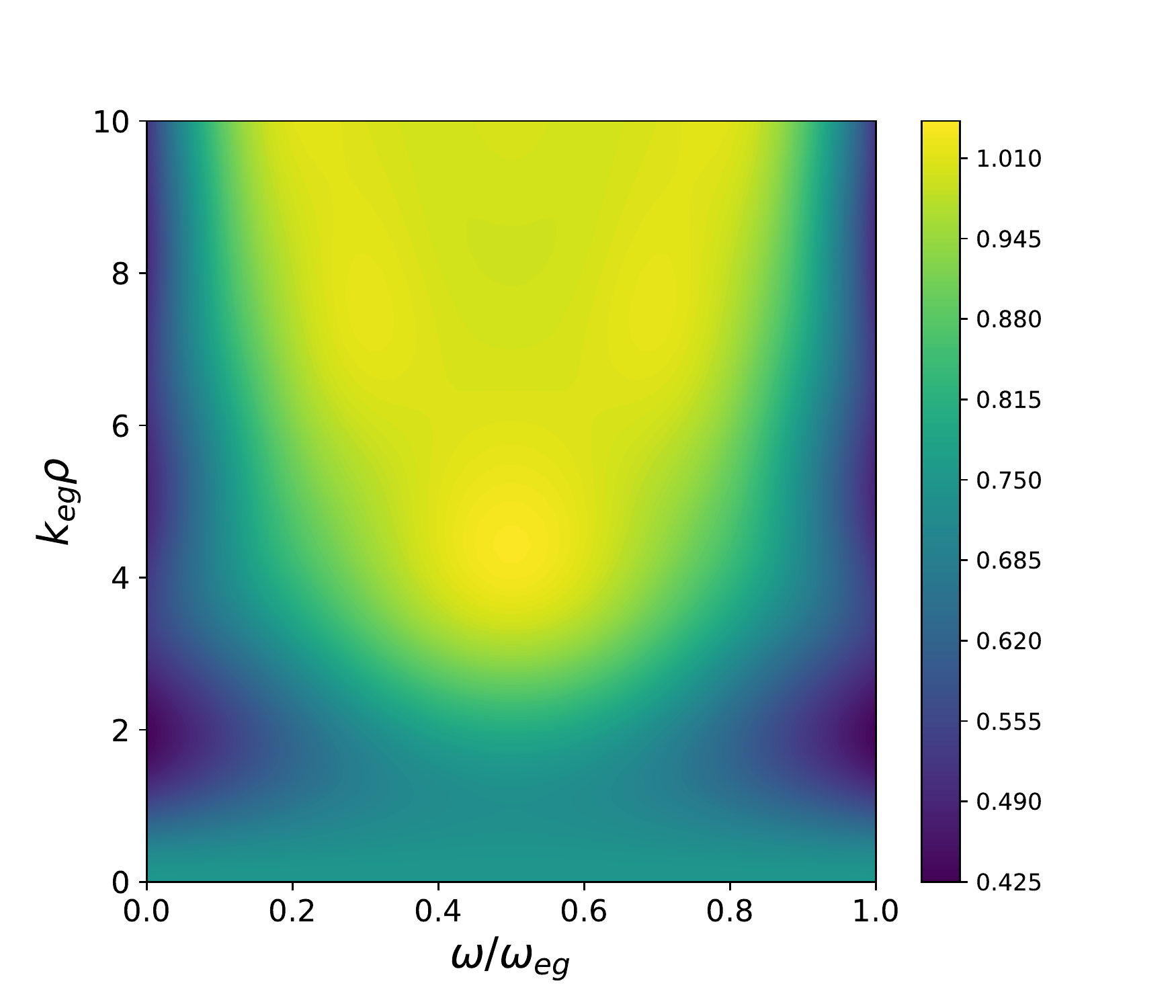}
        \caption{}
        \label{fig:contorno1}
    \end{subfigure}
    \\
    \begin{subfigure}[b]{\columnwidth}
        \centering
        \includegraphics[width=\textwidth]{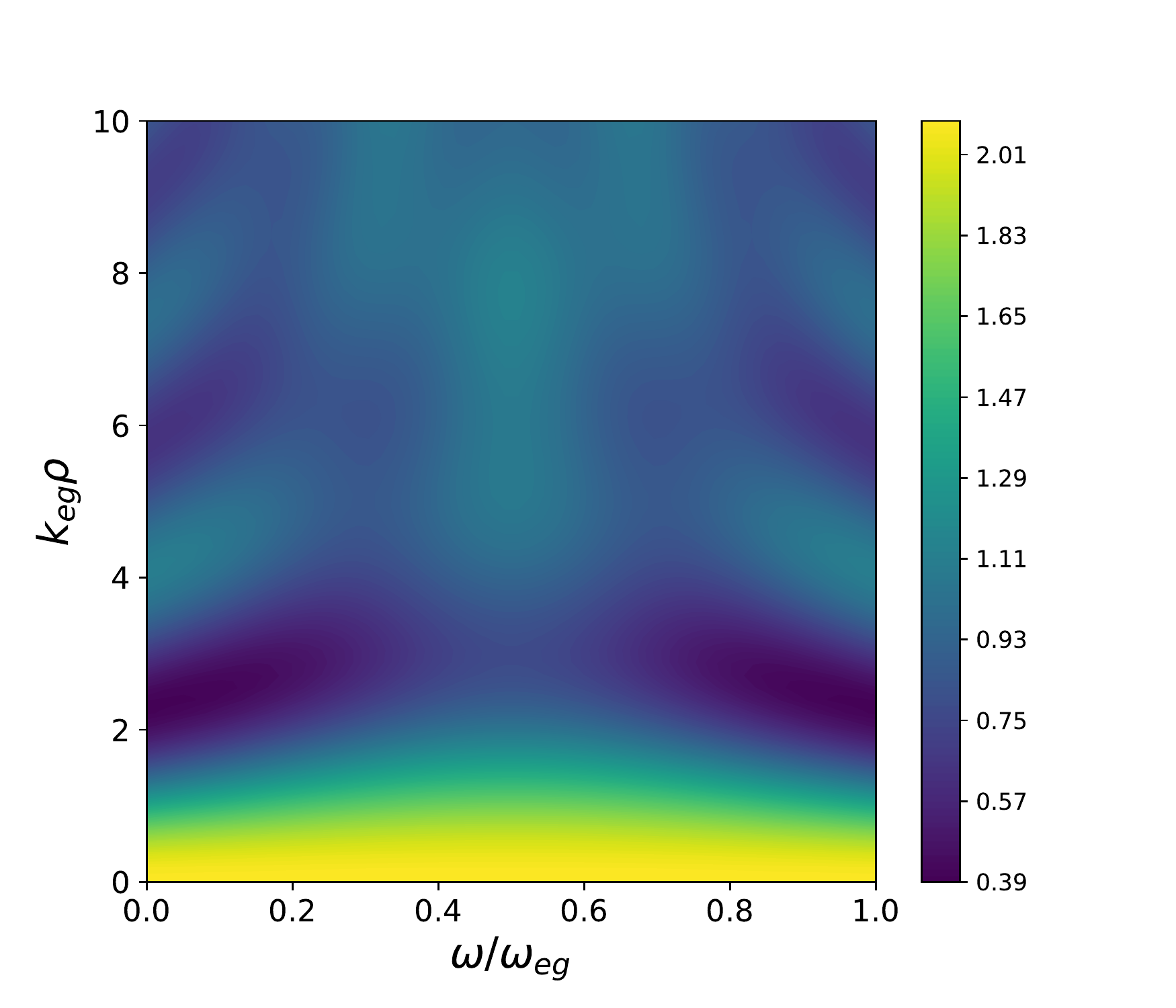}
        \caption{}
        \label{fig:contorno2}
    \end{subfigure}
    \caption{Contour plot of the spectral enhancement (color bar) as a function of $\omega/\omega_{eg}$ and $k_{eg}\rho$ for: (\subref{fig:contorno1}) $q=1.5$, (\subref{fig:contorno2}) $q=2.5$.}
    \label{fig:doisfotons_contorno}
\end{figure}

In Figs. \ref{fig:contorno1} and \ref{fig:contorno2},  contour plots  for the spectral enhancement as a function of both $k_{eg}\rho$ (vertical axis) and 
$\omega/\omega_{eg}$ (horizontal axis) are depicted for $q=1.5$ and $q=2.5$ respectively. These contour plots contain in a compact way most of the previous results. For instance, if we trace a horizontal line in Fig.  \ref{fig:contorno2} at $k_{eg} \rho = 4$ we will reproduce exactly the plot of the spectral enhancement  as a function of $\omega/\omega_{eg}$ represented in Fig.  \ref{fig:doisfotons_omega_q} by the red dotted  line. Analogously,  vertical lines in these contour plots will reproduce the curves for the spectral enhancement as a function of the normalized distance $k_{eg}\rho$ for fixed frequencies. Consider, for instance, the vertical line  given by $\omega/\omega_{eg} = 0.2$ in the contour plot of Fig. \ref{fig:contorno2}. This vertical line will reproduce exactly the plot of the spectral enhancement as a function of $k_{eg}\rho$ represented by the red dotted line of Fig. \ref{fig:q-2-5-por-distancia}. It is also worth mentioning that contour plots have the advantage to allow us to analyze the region in the parameter space in which a particular range of the spectral enhancement can be found.

Finally, In Fig. \ref{fig:k-10-por-q} we depict the spectral enhancement as a function of parameter $q$  for $k_{eg}=10$ and different (fixed) values of $\omega/\omega_{eg}$. First of all,  in analogy to what happens in the OPSE case, note the non-monotonic behavior  of $\gamma/\gamma_0$ for each frequency as $q$ is increased. Since the TPSE rates is proportional to a sum of Purcell factor products, this behavior was already expected. 
Another feature that must be emphasized is that the spectral enhancement exhibits for $q\gg 1$ a quadratic dependence on $q$. This can be understood mathematically if we recall the approximations in Eqs. \eqref{aprox1}, \eqref{aprox2} and \eqref{aprox3}, which imply that the Purcell factors are proportional to $q$ for $q\gg1$; consequently, in this situation, we can infer that $\gamma/\gamma_0\propto q^2$.

\section{\label{sec:conclusions} Final remarks and conclusions}

In this manuscript, we investigated the OPSE rate as well as the TPSE rate of a quantum emitter in the background of  a cosmic string. In the former case we started by analyzing some characteristics of the problem, as for example the oscillatory behavior of the OPSE rate as a function of the distance between the emitter and the string, as discussed in \cite{Cai:2015ioa}. However, we have also explored many another aspects of this setup. For instance, we analyzed separately the contributions to the OPSE rate of each component of the transition dipole moment of the atom. In contrast to an atom close to a perfectly conducting plate, the contribution that does not vanish as $k_{eg}\rho \rightarrow 0$ is that associated to the component of the transition dipole moment parallel to the string (recall that, for the perfectly conducting plate,  due to the boundary condition on the parallel component of the electric field, in this limit the only contribution that survives is the one associated to the component  of the transition dipole moment perpendicular to the conducting plate). 
Furthermore, we have also analyzed the behavior of the OPSE rates as a function of parameter $q$, which encodes the dependence on the linear density mass of the string. Interestingly, we found a non-monotonic behavior, but as discussed previously in the text, for large values of $q$ an approximate linear dependence of the spectral enhancement with respect to parameter $q$ shows up.

Concerning the TPSE of an atom near a cosmic string, which in contrast to the OPSE process is a broadband phenomenon, we started by showing how the spectral density (through the spectral enhancement) is affected by the presence of the string. This effect is usually referred to as Purcell effect for a TPSE process.  Since the photons now may be emitted in the frequency interval from zero to the transition frequency, and owing to the fact that other length scales  are involved, a TPSE process is much richer than a OPSE one (it has a larger parameter space). For convenience, we considered only $s \rightarrow s$ transitions. As a consequence, the TPSE rate can be straightforwardly obtained from the Purcell factors of OPSE processes. 
Using these facts, we explored the richness of TPSE processes and analyzed the spectral enhancement in a variety of situations. For instance, we showed that the spectral enhancement at fixed frequencies also exhibits an oscillatory behavior with the distance of the emitter to the string, tending to one as this distance tends to infinite. However, the oscillatory pattern is more irregular than int he case of a OPSE process. Regarding the dependence on $q$, we found that, as in the OPSE case, the spectral enhancement for fixed frequencies exhibits a non-monotonic behavior with respect this parameter, but in the limit $q\gg1$ this dependence becomes quadratic, which in its turn is a direct consequence of the fact that the spectral enhancement is calculated from a product of two Purcell factors. Our results were synthetized in the contour plots, which show the richness of the parameter space.

Though a TPSE process is of second order in perturbation theory, while a OPSE is a first order one, this process can become very important if for some reason (selection rules) the OPSE process is prohibited. Since TPSE processes are important in the investigation of the cosmos (as mentioned in the introduction), we think our results may be useful  somehow in the search for the existence of cosmic strings. Specifically speaking, the existence this topological defect would leave a signature in the spectral enhancement of a quantum emitter in a very specific way, therefore the analysis of the emitter's emission spectrum may be relevant for the identification of cosmic strings.

\begin{acknowledgments}
The authors  thank Reinaldo Melo e Souza, Patrícia Abrantes and Aram Saharian for enlightening  discussions. This work was partially supported by the Coordena\c{c}\~{a}o de Aperfei\c{c}oamento de Pessoal de N\'{\i}vel Superior - Brasil (CAPES) - Finance Code 001 and by Conselho Nacional de Desenvolvimento Cient\'{\i}fico e Tecnol\'{o}gico - CNPq, 310365/2018-0 (C.F.) and 309982/2018-9 (C.A.D.Z.). C.A.D.Z is also partially supported by Fundação Carlos Chagas Filho de Amparo à Pesquisa do Estado do Rio de Janeiro – FAPERJ under grant E-26/201.447/2021.
\end{acknowledgments}

\appendix

\section{\label{app:modes} Electromagnetic modes in a cosmic string background}

In this appendix,  we show how to obtain the electromagnetic field modes in a cosmic string background metric. Since this metric has a conical geometry, outside the cosmic string, the electromagnetic fields satisfy Maxwell's equations in the flat space-time, namely, 
\begin{align}
    \div{\vb{E}}(\vb{r},t)&=0, \label{eq:gauss}\\
    \div{\vb{B}}(\vb{r},t)&=0, \label{eq:monopolo} \\
    \curl{\vb{E}}(\vb{r},t)&=-{\pdv{\vb{B}}{t}}(\vb{r},t), \label{eq:faraday} \\        
    \curl{\vb{B}}(\vb{r},t)&=\frac{1}{c^2}{\pdv{\vb{E}}{t}} (\vb{r},t). \label{eq:amperemaxwell}
\end{align}
To find the electromagnetic fields that solve the above equations, it is convenient to write the electric and magnetic fields in the form
\begin{align}
    \vb E(\vb r, t)&=\left[\vb E_{\perp}(\vb r_\perp)+\vu{z} E_z(\vb r_{\perp})\right]e^{i(k_z z-\omega t)},\\
    \vb B(\vb r, t)&=\left[\vb B_{\perp}(\vb r_\perp)+\vu{z} B_z(\vb r_{\perp})\right]e^{i(k_z z-\omega t)},
\end{align}
where we decomposed the fields as a sum of  their perpendicular and parallel components to the $z$ axis, which coincides by assumption with the direction of the string. 
Following Ref. \cite{Zangwill}, plugging the above equations in Maxwell's Equations, one finds that the transverse electric (TE) and transverse magnetic (TM) fields are:
\begin{equation}\label{eq:ETEMEzBz}
    \begin{split}
        &\mathbf{E}_{\text{TE}}=-\frac{i\omega}{k_\perp^2 c}[\vu{z}\times\grad_\perp B_z]e^{i(k_z z-\omega t)},\\
        &\mathbf{E}_{\text{TM}}=\left[\frac{ik_z}{k_\perp^2}\grad_{\perp}E_z+E_z\vu{z}\right]e^{i(k_z z-\omega t)}.
    \end{split}
\end{equation}
where $\grad_\perp$ is the component of the gradient perpendicular to the string, $\grad_\perp=\grad-\vu{z} \partial_{z}$, and the z-components of fields satisfy the bi-dimensional Helmholtz equation
\begin{align}
    \left(\laplacian_\perp+k_\perp^2\right)F_z&=0 \label{eq:a8}
\end{align}
where $F_z=E_z,B_z$ and $\nabla_{\perp}^{2}=\grad_{\perp}\cdot\grad_{\perp}$. The latter equation can be solved by separation of variables imposing the cosmic string space-time boundary condition, namely, a periodicity of $\phi_0$ instead of $2\pi$, thus giving
\begin{equation}
    F_{z}=J_{q|m|}\left(k_\perp\rho\right)e^{iqm\phi}\;\;(m\in\mathbb{Z}),
\end{equation}
where $q=2\pi/\phi_{0}$. Now it suffices to substitute $E_{z}$ and $B_{z}$ into Eqs. \eqref{eq:ETEMEzBz} to find the solutions for the electric field
\begin{equation}
    \begin{split}
        &\mathbf{E}_{\text{TE}}=-\frac{i\omega}{k_\perp^2c}\vu{z}\times\left[\grad_\perp J_{q|m|}\left(k_\perp\rho\right)e^{i(qm\phi+k_z z-\omega t)}\right],\\
        &\mathbf{E}_{\text{TM}}=\left[\vu{z}+\frac{ik_z}{k_\perp^2}\grad_{\perp}\right]J_{q|m|}\left(k_\perp\rho\right)e^{i(qm\phi+k_z z-\omega t)}.
    \end{split}
\end{equation}
The vector potential, in the Coulomb gauge, can be obtained from the electric field from $\mathbf E=\partial_t \mathbf A$. Hence,
\begin{align}
    \label{eq:Ak0app}
    \vb{A}_{\vb k0}&=\frac{\beta_{\vb{k}0} c^2}{i\omega}\left(k_\perp^2\vu{z} + ik_z\grad_{\perp}\right)\\&\times\left[J_{q|m|}\left(k_\perp\rho\right)e^{i(qm\phi+k_z z-\omega t)}\right],\nonumber\\
    \vb{A}_{\vb{k}1}&=-\beta_{\vb{k}1} c \vu{z}\times\grad_{\perp}\left[J_{q|m|}\left(k_\perp\rho\right)e^{i(qm\phi+k_z z-\omega t)}\right],\label{eq:Ak1app}
\end{align}
where $\omega^2=(k_z^2+k^2_\perp)c^2$, $q=2\pi/\phi_0$, $m\in\mathbb{Z}$ and the indexes $0$ and $1$ stand for the polarizations of the modes and will be represented by $p$. Notice that we have introduced the constants  $\beta_{\vb{k}0}$ and $\beta_{\vb{k}1}$, which are necessary so that the vector potential modes obey the orthonormalization condition
\begin{equation}
    \label{eq:orto}
    \int \dd V \vb{A}_{\vb{k}p}\cdot\vb{A}^*_{\vb{k}'p'}=\delta^{3}(\vb{k}-\vb{k}')\delta_{p p'}.
\end{equation}
Performing the above integrals for $p=0, 1$, one finds
\begin{equation}
    \label{eq:normalizacao}
    |\beta_{\vb{k}0}|^2=|\beta_{\vb{k}1}|^2=\frac{q}{(2\pi k_\perp c)^2}.
\end{equation}


\section{\label{app:OPSErates} One-photon spontaneous emission rates computations}

Here, a detailed derivation of Eqs. \eqref{eq:Gammaz}, \eqref{eq:Gammarho} and \eqref{eq:Gammaphi} is presented. From the component of the electromagnetic modes which are parallel to the string, one finds
\begin{align}
    |\vb d^{\vu{z}}_{eg}\cdot\vb A_{\vb k  0}|^2&=|\vb d^{\vu{z}}_{eg}|^2\frac{q k_\perp^2c^3}{(2\pi\omega)^2}J^2_{q|m|}\left(k_\perp\rho\right),\\
    |\vb d^{\vu{z}}_{eg}\cdot\vb A_{\vb k  1}|^2&=0.
\end{align}
Inserting the above result into Eq. \eqref{eq:Gamma2}, we find that OPSE rate associated to the parallel component of the dipole parallel to the string reads
\begin{equation}
    \begin{split}
        \Gamma_{\vu{z}}&=\frac{\pi}{\epsilon_0\hbar}\sum_{\vb k}\,|\vb d^{\vu{z}}_{eg}\cdot\vb A_{\vb k  0}|^2\omega_k\delta(\omega_k-\omega_{eg})\\
        &=\frac{|\vb d^{\vu{z}}_{eg}|^2\pi}{\epsilon_0\hbar}\sum_{m=-\infty}^\infty \int \dd^2k\,\frac{q k_\perp^2c^2}{(2\pi\omega)^2}\omega_k\delta(\omega_k-\omega_{eg})J^2_{q|m|}\left(k_\perp\rho\right).
    \end{split}
\end{equation}
Using that $\omega^2=k^2c^2=(k_\perp^2+k_z^2)c^2$ e $\dd^2k = \dd k_\perp k_\perp \dd k_z$, one obtains
\begin{equation}\label{eq:GammazB4}
    \begin{split}
        \Gamma_{\vu{z}}&=\frac{|\vb d^{\vu{z}}_{eg}|^2\pi c}{(2\pi)^4\epsilon_0\hbar}\sum_{m=-\infty}^\infty \int_0^\infty \dd k_\perp\int_{-\infty}^{\infty} \dd k_z\,\frac{k_\perp^3}{\sqrt{k_\perp^2+k_z^2}}\\
        &\times\delta\left(\sqrt{k_\perp^2c^2+k_z^2c^2}-\omega_{eg}\right)J^2_{q|m|}\left(k_\perp\rho\right)\\
    \end{split}
\end{equation}
From the following property of the delta function
\begin{equation}
    \delta(f(x)) = \sum_i\frac{\delta(x-x_i)}{|f'(x_i)|}
\end{equation}
where $x_i$ are the zeroes of $f(x)$, and considering first an integration in $k_z$, we can expand the delta function as
\begin{equation}
    \label{eq:delta}
    \begin{split}
        \delta(\omega(k_\perp, k_z)-\omega_{eg})&=\frac{\omega_{eg}}{c\sqrt{\omega^2_{eg}-k_\perp^2c^2}}\\
        &\left[\delta\left(k_z-\frac1c\sqrt{\omega^2_{eg}-k_\perp^2c^2}\right)\right.\\
        &\left.+\delta\left(k_z+\frac1c\sqrt{\omega^2_{eg}-k_\perp^2c^2}\right)\right].
    \end{split}
\end{equation}
Substituting the above equation into Eq. \eqref{eq:GammazB4}, we find
\begin{equation}
    \begin{split}
        \Gamma_{\vu{z}}
        &=\frac{|\vb d^{\vu{z}}_{eg}|^2 q\omega^3_{eg}}{2\pi\epsilon_0\hbar c^3}\sum_{m=-\infty}^\infty \int_0^1 \dd u\,\frac{u^3}{\sqrt{1-u^2}}J^2_{q|m|}\left(k_{eg}\rho u\right),
    \end{split}
\end{equation}
where there was made the redefinition of variables $u=k_\perp c/\omega_{eg}$ and $k_{eg}= \omega_{eg}/c$. Finally, by comparing it with the free-space rate $\Gamma_{0}$ given in Eq. \eqref{eq:Gamma0}, we are left with
\begin{equation}
    \frac{\Gamma_{\vu z}}{\Gamma_{0}}=\frac{|\vb d_{eg}^{\vu z}|^2}{|\vb d_{eg}|^2}\frac{3q}{2}\sum_{m=-\infty}^\infty \int_0^1 \dd u\,\frac{u^3}{\sqrt{1-u^2}}J^2_{q|m|}\left(k_{eg}\rho u\right)
\end{equation}

Now we compute the OPSE rates associated to the perpendicular components of the dipole with respect to the string, namely, the radial and tangential ones. Using Eqs. \eqref{eq:Amodo0}, \eqref{eq:Amodo1} and \eqref{eq:normalizacao}, one obtains:
\begin{align}
    |\vb d^{\vu*\rho}_{eg}\cdot\vb A_{\vb k  0}|^2&=|\vb d^{\vu*\rho}_{eg}|^2\frac{q k_{z}^{2}c^{2}}{(2\pi\omega)^2}\left[{J'_{q|m|}}\left(k_\perp\rho\right)\right]^{2}, \\
    |\vb d^{\vu*\rho}_{eg}\cdot\vb A_{\vb k  1}|^2&=|\vb d^{\vu*\rho}_{eg}|^2\frac{q^{3}m^{2}}{(2\pi)^{2}k_\perp^2\rho^{2}}J^2_{q|m|}\left(k_\perp\rho\right).\\
    |\vb d^{\vu*\phi}_{eg}\cdot\vb A_{\vb k  0}|^2&=|\vb d^{\vu*\phi}_{eg}|^2\frac{q^{3}k_{z}^{2}c^{2}m^{2}}{(2\pi)^{2}k_\perp^2\omega^{2}\rho^{2}}J^2_{q|m|}\left(k_\perp\rho\right), \\
    |\vb d^{\vu*\phi}_{eg}\cdot\vb A_{\vb k  1}|^2&=|\vb d^{\vu*\phi}_{eg}|^2\frac{q }{(2\pi)^2}\left[{J'_{q|m|}}\left(k_\perp\rho\right)\right]^{2}.
\end{align}
where the prime indicates the derivative with respect to the argument of the function. Summing over the polarizations and using the following Bessel function identities \cite{abramowitz},
\begin{align}
    &\dv{J_\nu(x)}{x}=\frac{J_{\nu-1}(x)-J_{\nu+1}(x)}{2},\\
    &\frac{\nu}x J_\nu(x)=\frac{J_{\nu-1}(x)+J_{\nu+1}(x)}{2},
\end{align}
we have, for the radial component
\begin{align}
    &\sum_p |\vb d^{\vu*\rho}_{eg}\cdot\vb A_{\vb k  p}|^2=\frac{|\vb d^{\vu*\rho}_{eg}|^2q}{4(2\pi)^2}\left\{\left(1+\frac{k_z^2c^2}{\omega^2}\right)\left[J^2_{q|m|-1}\left(k_\perp\rho\right)+\right.\right.\nonumber \\
    &\left.\left.+J^2_{q|m|+1}\left(k_\perp\rho\right)\right]+\frac{2k_\perp^2c^2}{\omega^2}J_{q|m|-1}\left(k_\perp\rho\right)J_{q|m|+1}\left(k_\perp\rho\right)\right\};
\end{align}
whereas for the tangential component
\begin{align}
    &\sum_p |\vb d^{\vu*\phi}_{eg}\cdot\vb A_{\vb k  p}|^2=\frac{|\vb d^{\vu*\phi}_{eg}|^2q}{4(2\pi)^2}\left\{\left(1+\frac{k_z^2c^2}{\omega^2}\right)\left[J^2_{q|m|-1}\left(k_\perp\rho\right)+\right.\right.\nonumber \\
    &\left.\left.+J^2_{q|m|+1}\left(k_\perp\rho\right)\right]-\frac{2k_\perp^2c^2}{\omega^2}J_{q|m|-1}\left(k_\perp\rho\right)J_{q|m|+1}\left(k_\perp\rho\right)\right\}.
\end{align}
Lastly, inserting the above expressions into Eq. \eqref{eq:Gamma2} and normalizing the result with respect to the free-space rate, we find
\begin{align}
    \label{eq:Gammarho2}
    \frac{\Gamma_{\vu*\rho}}{\Gamma_0}=&\frac{|\vb d_{eg}^{\vu*\rho}|^2}{|\vb d_{eg}|^2}\frac{3q}{8}\sum_{m=-\infty}^\infty \int_0^{1} \dd u\,\frac{u}{\sqrt{1-u^2}}\nonumber\\& \left[\left(2-u^2\right)\left(J^2_{q|m|-1}\left(k_{eg}\rho u\right)+J^2_{q|m|+1}\left(k_{eg}\rho u\right)\right)\right.\nonumber\\&+2u^2J_{q|m|-1}\left(k_{eg}\rho u\right)J_{q|m|+1}\left(k_{eg}\rho u\right)\Big]
\end{align}
\begin{align}
    \label{eq:Gammaphi2}
    \frac{\Gamma_{\vu*\phi}}{\Gamma_0}&=\frac{|\vb d_{eg}^{\vu*\phi}|^2}{|\vb d_{eg}|^2}\frac{3q}{8}\sum_{m=-\infty}^\infty \int_0^{1} \dd u\,\frac{u}{\sqrt{1-u^2}}\nonumber \\
    &\left[\left(2-u^2\right)\left(J^2_{q|m|-1}\left(k_{eg}\rho u\right)+J^2_{q|m|+1}\left(k_{eg}\rho u\right)\right)\right.\nonumber\\
    &\left.-2u^2J_{q|m|-1}\left(k_{eg}\rho u\right)J_{q|m|+1}\left(k_{eg}\rho u\right)\right]
\end{align}
where we made the same change of variables in the integration as in the parallel component case, namely, $u=k_\perp c/\omega_{eg}$ and $k_{eg}= \omega_{eg}/c$.

\bibliography{bibliography}

\providecommand{\noopsort}[1]{}\providecommand{\singleletter}[1]{#1}%
\begin{thebibliography}{10}

\bibitem{milonni1984}
Peter~W Milonni.
\newblock Why spontaneous emission?
\newblock {\em American Journal of Physics}, 52(4):340--343, 1984.

\bibitem{purcell1951}
HI~Ewen and EM~Purcell.
\newblock Radiation from galactic hydrogen at 1,420 {M}c/sec.
\newblock {\em Nature}, 168(356):115--125, 1951.

\bibitem{furlanetto2006}
Steven~R Furlanetto, S~Peng Oh, and Frank~H Briggs.
\newblock Cosmology at low frequencies: The 21 cm transition and the
  high-redshift universe.
\newblock {\em Physics reports}, 433(4-6):181--301, 2006.

\bibitem{pritchard2012}
Jonathan~R Pritchard and Abraham Loeb.
\newblock 21 cm cosmology in the 21st century.
\newblock {\em Reports on Progress in Physics}, 75(8):086901, 2012.

\bibitem{breit1940}
G~Breit and E~Teller.
\newblock Metastability of {H}ydrogen and {H}elium levels.
\newblock {\em The Astrophysical Journal}, 91:215, 1940.

\bibitem{spitzer1951}
Lyman Spitzer~Jr and Jesse~L Greenstein.
\newblock Continuous emission from planetary nebulae.
\newblock {\em The Astrophysical Journal}, 114:407, 1951.

\bibitem{gurzadyan2013}
Grigor~A Gurzadyan.
\newblock {\em The physics and dynamics of planetary nebulae}.
\newblock Springer Science \& Business Media, 2013.

\bibitem{wong2006}
Wan~Yan Wong, Sara Seager, and Douglas Scott.
\newblock Spectral distortions to the cosmic microwave background from the
  recombination of hydrogen and helium.
\newblock {\em Monthly Notices of the Royal Astronomical Society},
  367(4):1666--1676, 2006.

\bibitem{hirata2008}
Christopher~M Hirata.
\newblock Two-photon transitions in primordial hydrogen recombination.
\newblock {\em Physical Review D}, 78(2):023001, 2008.

\bibitem{chluba2011}
J~Chluba and RM~Thomas.
\newblock Towards a complete treatment of the cosmological recombination
  problem.
\newblock {\em Monthly Notices of the Royal Astronomical Society},
  412(2):748--764, 2011.

\bibitem{ilakovac2006two}
Ksenofont Ilakovac, Milivoj Uroi{\'c}, Marija Majer, Selim Pa{\v{s}}i{\'c}, and
  Branko Vukovi{\'c}.
\newblock Two-photon decay of k-shell vacancy states in heavy atoms.
\newblock {\em Radiation Physics and Chemistry}, 75(11):1451--1460, 2006.

\bibitem{purcell1946}
E.~M. Purcell.
\newblock Spontaneous emission probabilities at radio frequencies.
\newblock {\em Physical review}, 69:681, 1946.

\bibitem{Haroche}
Serge Haroche.
\newblock Nobel lecture: Controlling photons in a box and exploring the quantum
  to classical boundary.
\newblock {\em Rev. Mod. Phys.}, 85:1083--1102, Jul 2013.

\bibitem{Lodahl}
Peter Lodahl, Sahand Mahmoodian, and S{\o}ren Stobbe.
\newblock Interfacing single photons and single quantum dots with photonic
  nanostructures.
\newblock {\em Reviews of Modern Physics}, 87(2):347, 2015.

\bibitem{rivera2017making}
Nicholas Rivera, Gilles Rosolen, John~D Joannopoulos, Ido Kaminer, and Marin
  Solja{\v{c}}i{\'c}.
\newblock Making two-photon processes dominate one-photon processes using
  mid-{IR} phonon polaritons.
\newblock {\em Proceedings of the National Academy of Sciences},
  114(52):13607--13612, 2017.

\bibitem{rivera2016shrinking}
Nicholas Rivera, Ido Kaminer, Bo~Zhen, John~D Joannopoulos, and Marin
  Solja{\v{c}}i{\'c}.
\newblock Shrinking light to allow forbidden transitions on the atomic scale.
\newblock {\em Science}, 353(6296):263--269, 2016.

\bibitem{muniz2020two}
Y~Muniz, A~Manjavacas, C~Farina, DAR Dalvit, and WJM Kort-Kamp.
\newblock Two-photon spontaneous emission in atomically thin plasmonic
  nanostructures.
\newblock {\em Physical Review Letters}, 125(3):033601, 2020.

\bibitem{Bekenstein:1977mv}
J.D. Bekenstein and A.~Meisels.
\newblock {Einstein a and B Coefficients for a Black Hole}.
\newblock {\em Phys. Rev. D}, 15:2775--2781, 1977.

\bibitem{Cai:2015ioa}
Huabing Cai, Hongwei Yu, and Wenting Zhou.
\newblock {Spontaneous excitation of a static atom in a thermal bath in cosmic
  string spacetime}.
\newblock {\em Phys. Rev. D}, 92(8):084062, 2015.

\bibitem{Mermin:1979zz}
N.D. Mermin.
\newblock {The topological theory of defects in ordered media}.
\newblock {\em Rev. Mod. Phys.}, 51:591--648, 1979.

\bibitem{vilenkin2000cosmic}
Alexander Vilenkin and E~Paul~S Shellard.
\newblock {\em Cosmic strings and other topological defects}.
\newblock Cambridge University Press, Cambridge, UK, 2000.

\bibitem{Vachaspati:2015cma}
Tanmay Vachaspati, Levon Pogosian, and Daniele Steer.
\newblock {Cosmic Strings}.
\newblock {\em Scholarpedia}, 10(2):31682, 2015.

\bibitem{ellis2020cosmic}
John Ellis and Marek Lewicki.
\newblock Cosmic {S}tring {I}nterpretation of {NANOG}rav {P}ulsar timing data.
\newblock {\em arXiv preprint arXiv:2009.06555}, 2020.

\bibitem{blasi2020has}
Simone Blasi, Vedran Brdar, and Kai Schmitz.
\newblock Has {NANOG}rav found first evidence for cosmic strings?
\newblock {\em arXiv preprint arXiv:2009.06607}, 2020.

\bibitem{samanta2021gravitational}
Rome Samanta and Satyabrata Datta.
\newblock Gravitational wave complementarity and impact of nanograv data on
  gravitational leptogenesis.
\newblock {\em Journal of High Energy Physics}, 2021(5):1--24, 2021.

\bibitem{Moraes:2000xa}
F.~Moraes.
\newblock {Condensed matter physics as a laboratory for gravitation and
  cosmology}.
\newblock {\em Braz. J. Phys.}, 30:304--308, 2000.

\bibitem{Hu:2017ytf}
Jiawei Hu and Hongwei Yu.
\newblock {Manipulating lightcone fluctuations in an analogue cosmic string}.
\newblock {\em Phys. Lett. B}, 777:346--350, 2018.

\bibitem{Vilenkin:1981kz}
A.~Vilenkin.
\newblock {Cosmic Strings}.
\newblock {\em Phys. Rev. D}, 24:2082--2089, 1981.

\bibitem{Hiscock:1985uc}
W.~A. Hiscock.
\newblock {Exact Gravitational Field of a String}.
\newblock {\em Phys. Rev. D}, 31:3288--3290, 1985.

\bibitem{vilenkin1984cosmic}
Alexander Vilenkin.
\newblock Cosmic strings as gravitational lenses.
\newblock {\em The Astrophysical Journal}, 282:L51--L53, 1984.

\bibitem{linet}
B.~Linet.
\newblock Force on a charge in the space-time of a cosmic string.
\newblock {\em Phys. Rev. D}, 33:1833--1834, Mar 1986.

\bibitem{aliev1989gravitational}
A.~N. Aliev and D.~V. Gal'Tsov.
\newblock Gravitational {A}haronov-{B}ohm radiation in string-generated conical
  space-time.
\newblock {\em Annals of Physics}, 193(1):142--165, 1989.

\bibitem{Milonni}
P.W. Milonni.
\newblock {\em {The Quantum vacuum: An Introduction to quantum
  electrodynamics}}.
\newblock {Academic Press}, {San Diego}, 1994.

\bibitem{Dirac:1927}
P.~A.~M. Dirac.
\newblock {The quantum theory of emission and absorption of radiation}.
\newblock {\em Proc. R. Soc. Lond. A}, 114:243--265, 1927.

\bibitem{abramowitz}
M.~Abramowitz and I.~A. Stegun.
\newblock {\em Handbook of Mathematical Functions: With Formulas, Graphs, and
  Mathematical Tables}.
\newblock Applied mathematics series. Dover Publications, 1965.

\bibitem{muniz2019}
Y.~Muniz, F.~S.~S. da~Rosa, C.~Farina, D.~Szilard, and W.~J.~M. Kort-Kamp.
\newblock Quantum two-photon emission in a photonic cavity.
\newblock {\em Physical Review A}, 100(2):023818, 2019.

\bibitem{Saharian:2011sx}
A.A. Saharian and A.S. Kotanjyan.
\newblock {Repulsive Casimir-Polder forces from cosmic strings}.
\newblock {\em Eur. Phys. J. C}, 71:1765, 2011.

\bibitem{Mayer}
Maria Göppert‐Mayer.
\newblock Über elementarakte mit zwei quantensprüngen.
\newblock {\em Annalen der Physik}, 401(3):273--294, 1931.

\bibitem{Novotny}
L.~Novotny and B.~Hecht.
\newblock {\em {Principles of Nano-optics}}.
\newblock Cambridge Univ. Press, Cambridge, UK, 2012.

\bibitem{Zangwill}
A.~Zangwill.
\newblock {\em {Modern electrodynamics}}.
\newblock Cambridge Univ. Press, Cambridge, UK, 2013.

\end{thebibliography}

\bibliographystyle{unsrt}

\end{document}